\documentclass[aps,onecolumn,tightenlines,showpacs,nofootinbib,preprintnumbers,
amsmath,amssymb]{revtex4}

\usepackage[usenames,dvipsnames]{color}
\usepackage{dcolumn}
\usepackage{graphicx}
\usepackage{subfigure}
\usepackage{epstopdf}
\usepackage{bm}
\usepackage{lscape}
\renewcommand{\arraystretch}{0.9}

\usepackage{color}
\usepackage[colorlinks,
linkcolor=blue,
filecolor=blue,
anchorcolor=blue,
urlcolor=blue,
citecolor=blue,
bookmarks=false,
]{hyperref}
\usepackage{url}

\usepackage{setspace}

\begin{document}
\begin{spacing}{1.5}

\title{Hunting for direct CP violation in $\bar{B}_s^0 \to {\pi^+}{\pi^-}K^{*0}$}

\author{
Sheng-Tao Li$^{1}$\footnote{lst@mails.ccnu.edu.cn}, Gang L\"{u}$^{2}$\footnote{ganglv66@sina.com}
}\label{key}

\affiliation{
\small $^{1}$Institute of Particle Physics and Key Laboratory of Quark and Lepton Physics~(MOE), \\
\small Central China Normal University, Wuhan, Hubei 430079, China  \\
\small $^{2}$College of Science, Henan University of Technology, Zhengzhou 450001, China
}

\begin{abstract}
In perturbative QCD approach, based on the first order of isospin symmetry breaking, we study the direct $CP$ violation in the decay of $\bar{B}_s^0\to\rho(\omega )K^{*0}\to {\pi^+}{\pi^-}K^{*0}$.
An interesting mechanism is applied to enlarge the $CP$ violating asymmetry involving the charge symmetry breaking between $\rho$ and $\omega$. We find that the $CP$ violation is large by the $\rho-\omega$ mixing mechanism when the invariant masses of the $\pi^+\pi^-$ pairs is in the vicinity of the $\omega$ resonance. For the decay process of $\bar{B}^0_{s}\to\rho(\omega )K^{*0}\to{\pi^+}{\pi^-}K^{*0}$, the maximum $CP$ violation can reach $-59.12\%$. Furthermore, taking $\rho-\omega$ mixing into account, we calculate the branching ratio for $\bar{B}_s^0 \rightarrow \rho(\omega) K^{*0}$. We also discuss the possibility of observing the predicted $CP$ violation asymmetry at the LHC.
\end{abstract}


\maketitle

\section{\label{intro}Introduction}

Charge-Parity ($CP$) violation is an open problem, even though it has been known in the Neutral kaon systems for more than five decades \cite{Christenson:1964fg}. The study of $CP$ violation in the heavy quark systems is important to our understanding of both particle physics and the evolution of the early universe. Within the standard model (SM), $CP$ violation is related to the non-zero weak complex phase angle from the  Cabibbo-Kobayashi-Maskawa (CKM) matrix, which describes the mixing of the three generations of quarks \cite{Cabibbo:1963yz,Kobayashi:1973fv}. Theoretical studies predicted large $CP$ violation in the $B$ meson system \cite{Carter:1980hr,Carter:1980tk,Bigi:1981qs}. In recent years, the LHCb collaboration has measured sizable direct CP asymmetries in the phase space of the three-body decay channels of $B^{\pm}\rightarrow \pi^{\pm}\pi^{+}\pi^{-}$ and $B^{\pm}\rightarrow K^{\pm}\pi^{+}\pi^{-}$ \cite{Aaij:2013bla,Aaij:2014iva,Bediaga:2020qxg}. These processes are also valuable for studying the mechanism of multi-body heavy meson decays. Hence, more attention has been focused on the non-leptonic $B$ meson three-body decays channels in searching for $CP$ violation, both theoretically and experimentally.

Direct $CP$ violation in $b$ hadron decays occurs through the interference of at least two amplitudes with different weak phase $\phi$ and strong phase $\delta$. The weak phase difference $\phi$ is determined by the CKM matrix elements, while the strong phase can be produced by the hadronic matrix elements and interference between the intermediate states. The hadronic matrix elements are not still well determined by the theoretical approach. The mechanism of two-body $B$ decay is still not quite clear, although many physicists are devoted to this field. Many factorization approaches have been developed to calculate the two-body hadronic decays, such as the naive factorization approach \cite{Fakirov:1977ta,Cabibbo:1977zv,Wirbel:1985ji,Bauer:1986bm}, the QCD factorization (QCDF) \cite{Beneke:1999br,Beneke:2000ry,Sachrajda:2001uv,Beneke:2001ev,Beneke:2003zv}, perturbative QCD (pQCD) \cite{Lu:2000em,Keum:2000ph,Keum:2000wi}, and soft-collinear effective theory (SCET) \cite{Bauer:2000yr,Bauer:2001cu,Bauer:2001yt}. Most factorization approaches are based on heavy quark expansion and light-cone expansion in which only the leading power or part of the next to leading power contributions are calculated to compare with the experiments. However, the different methods may present different strong phases so as to affect the value of the $CP$ violation. Meanwhile, in order to have a large signal of $CP$ violation, we need appeal to some phenomenological mechanism to obtain a large strong phase $\delta$. In Refs. \cite{Enomoto:1996cv,Guo:1998eg,Guo:2000uc,Leitner:2002xh,Lu:2010vb,Lu:2011zzf}, the authors studied direct $CP$ violation in hadronic $B$ (include $B_s$ and $\Lambda_b$) decays through the interference of tree and penguin diagrams, where $\rho$-$\omega$ mixing was used for this purpose in the past few years and focused on the naive factorization and QCD factorization approaches. This mechanism was also applied to generalize the pQCD approach to the three-body non-leptonic decays in ${B^{0, \pm }} \to {\pi ^{0, \pm }}{\pi ^ + }{\pi ^ - }$ and $B_{c}\rightarrow D_{(s)}^+\pi^+\pi^-$ where even larger $CP$ violation may be possible \cite{Lu:2013jma,Lu:2016lgc}. In this paper, we will investigate direct $CP$ violation of the decay process $\bar{B}^0_{s}\to\rho(\omega )K^{*0}\to{\pi^+}{\pi^-}K^{*0}$ involving the same mechanism in the pQCD approach.

Three-body decays of heavy $B$ mesons are more complicated than the two-body decays as they receive both resonant and non-resonant contributions. Unlike the two-body case, to date we still do not have effective theories for hadronic three-body decays, though attempts along the framework of pQCD and QCDF have been used in the past \cite{Krankl:2015fha,Wang:2014ira,Chen:2002th,Qi:2018syl}. As a working starting point, we intend to study $\rho$-$\omega$ mixing effect in three-body decays of the $B$ meson. The $\rho$-$\omega$ mixing mechanism is caused by the isospin symmetry breaking from the mixing between the $u$ and $d$ flavors \cite{Fritzsch:2000pg,Fritzsch:2001aj}. In Ref. \cite{OConnell:1995nse}, the authors studied  the $\rho-\omega$ mixing and the pion form factor in the time-like region, where $\rho-\omega$ mixing comes from three part contributions: two from the direct coupling of the quasi-two-body decay of $\bar{B}^0_{s}\to \rho K^{*0} \to \pi^+\pi^-K^{*0}$ and $\bar{B}^0_{s}\to\omega K^{*0}\to \pi^+\pi^-K^{*0}$ and the other from the interference of $\bar{B}^0_{s}\to\omega K^{*0} \to \rho K^{*0}  \to \pi^+\pi^-K^{*0}$ mixing. Generally speaking, the amplitudes of their contributions: $\bar{B}^0_{s}\to \rho K^{*0} \to \pi^+\pi^-K^{*0}$ $>$ $\bar{B}^0_{s}\to\omega K^{*0} \to \rho K^{*0}  \to \pi^+\pi^-K^{*0}$ $>$ $\bar{B}^0_{s}\to\omega K^{*0}\to \pi^+\pi^-K^{*0}$. $\omega \to \pi^+\pi^-$ and $\omega \to \rho  \to \pi^+\pi^-$ were used to obtain the (effective)  mixing  matrix element $\widetilde{\Pi}_{\rho\omega}(s)$  \cite{Bernicha:1994re,Maltman:1996kj,OConnell:1997ggd}. The magnitude has been determined by the pion form factor through the data from the cross section of ${e^+}{e^-} \to {\pi^+}{\pi^-}$ in the $\rho$ and $\omega$ resonance region \cite{OConnell:1995nse,OConnell:1996amv,OConnell:1997ggd,Wolfe:2009ts,Wolfe:2010gf}. Recently, isospin symmetry breaking was discussed by incorporating the vector meson dominance (VMD) model in the weak decay process of the  meson \cite{Gardner:1997yx,Guo:1999ip,Guo:2000uc,Lu:2016lgc,Lu:2014uja,Li:2019xwh}. However, one can find that $\rho-\omega$ mixing produces the large $CP$ violation from the effect of isospin symmetry breaking in the three and four bodies decay process. Hence, in this paper, we shall follow the method of Refs. \cite{Gardner:1997yx,Guo:1999ip,Guo:2000uc,Lu:2016lgc,Lu:2014uja,Li:2019xwh} to investigate the decay process of $\bar{B}^0_{s}\to\rho(\omega )K^{*0}\to{\pi^+}{\pi^-}K^{*0}$ by isospin symmetry breaking.

The remainder of this paper is organized as follows. In Sec. \ref{sec:hamckm} we will present the form of the effective Hamiltonian and briefly introduce the pQCD framework and wave functions. In Sec. \ref{sec:cpv1} we give the calculating formalism and details of the $CP$ violation from $\rho-\omega$ mixing
in the decay process $\bar{B}_s^0\to\rho(\omega )K^{*0}\to {\pi^+}{\pi^-}K^{*0}$. In Sec. \ref{Br} we calculate the branching ratio for decay process of $\bar{B}^0_{s}\rightarrow \rho^{0}(\omega)K^{*0}$. In Sec. \ref{int} we show the input parameters. We present the numerical results in Sec. \ref{num}.
Summary and discussion are included in
Sec. \ref{sec:conclusion}. The related function defined in the text are given
in Appendix.

\section{\label{sec:hamckm}The FrameWork}

Based on the operator product expansion, the effective weak Hamiltonian for the decay processes $\bar B_s^0\to \rho^{0}(\omega)K^{*0}$ can be expressed as \cite{Buchalla:1995vs}
\begin{eqnarray}
 {\cal H}_{eff} &=& \frac{G_{F}}{\sqrt{2}}
     \Bigg\{ V_{ub} V_{ud}^{\ast} \Big[
     C_{1}({\mu}) Q^{u}_{1}({\mu})
  +  C_{2}({\mu}) Q^{u}_{2}({\mu})\Big]
  -V_{tb} V_{td}^{\ast} \Big[{\sum\limits_{i=3}^{10}} C_{i}({\mu}) Q_{i}({\mu})
  \Big ] \Bigg\} + \mbox{H.c.} ,
 \label{2a}
 \vspace{2mm}
 \end{eqnarray}
where $G_F$ represents the Fermi constant, $C_i(\mu)$ (i=1,...,10) are the Wilson coefficients, and 
$V_{ub}$, $V_{ud}$, $V_{tb}$, and $V_{td}$ are the CKM matrix element. The
operators $O_i$ have the following forms:
\begin{eqnarray}
O^{u}_1&=& \bar d_\alpha \gamma_\mu(1-\gamma_5)u_\beta\bar
u_\beta\gamma^\mu(1-\gamma_5)b_\alpha,\nonumber\\
O^{u}_2&=& \bar d \gamma_\mu(1-\gamma_5)u\bar
u\gamma^\mu(1-\gamma_5)b,\nonumber\\
O_3&=& \bar d \gamma_\mu(1-\gamma_5)b \sum_{q'}
\bar q' \gamma^\mu(1-\gamma_5) q',\nonumber\\
O_4 &=& \bar d_\alpha \gamma_\mu(1-\gamma_5)b_\beta \sum_{q'}
\bar q'_\beta \gamma^\mu(1-\gamma_5) q'_\alpha,\nonumber\\
O_5&=&\bar d \gamma_\mu(1-\gamma_5)b \sum_{q'} \bar q'
\gamma^\mu(1+\gamma_5)q',\nonumber\\
O_6& = &\bar d_\alpha \gamma_\mu(1-\gamma_5)b_\beta \sum_{q'}
\bar q'_\beta \gamma^\mu(1+\gamma_5) q'_\alpha,\nonumber\\
O_7&=& \frac{3}{2}\bar d \gamma_\mu(1-\gamma_5)b \sum_{q'}
e_{q'}\bar q' \gamma^\mu(1+\gamma_5) q',\nonumber\\
O_8 &=&\frac{3}{2} \bar d_\alpha \gamma_\mu(1-\gamma_5)b_\beta \sum_{q'}
e_{q'}\bar q'_\beta \gamma^\mu(1+\gamma_5) q'_\alpha,\nonumber\\
O_9&=&\frac{3}{2}\bar d \gamma_\mu(1-\gamma_5)b \sum_{q'} e_{q'}\bar q'
\gamma^\mu(1-\gamma_5)q',\nonumber\\
O_{10}& = &\frac{3}{2}\bar d_\alpha \gamma_\mu(1-\gamma_5)b_\beta \sum_{q'}
e_{q'}\bar q'_\beta \gamma^\mu(1-\gamma_5) q'_\alpha,
\label{2b}
\vspace{2mm}
\end{eqnarray}
where $\alpha$ and $\beta$ are SU(3) color indices, $e_{q^\prime}$ is the electric charge of quark $q^\prime$ in the unit of $|e|$, and the sum extend over  $q^\prime=u, d, s, c$ or $b$ quarks. In Eq. (\ref{2b}) $O_1^u$ and $O_2^u$ are tree
operators, $O_3$--$O_6$ are QCD penguin operators and $O_7$--$O_{10}$ are
the operators associated with electroweak penguin diagrams.

The Wilson coefficient $C_{i}(\mu)$ in Eq. (\ref{2a}) describes the coupling strength for a given operator and summarizes the physical contributions from scales higher than $\mu$~\cite{Buras:1998raa}. They are calculable perturbatively with the renormalization group improved perturbation theory. Usually, the scale $\mu$ is
chosen to be of order ${\cal O}(m_{b})$ for $B$ meson decays. Since we work in the leading order of perturbative QCD
($O(\alpha_s )$), it is consistent to use the leading order Wilson
coefficients. So, we use numerical values of $C_i(m_b)$ as follow \cite{Keum:2000wi,Lu:2000em}:
\begin{eqnarray}
C_1 &=&-0.2703,\;\; \; \; \; \; \,C_2=1.1188,\nonumber\\
C_3 &=& 0.0126,\;\; \; \; \; \; \; \; \; \;C_4 = -0.0270,\nonumber\\
C_5 &=& 0.0085,\;\; \; \; \; \; \; \; \; \;C_6 = -0.0326,\nonumber\\
C_7 &=& 0.0011,\;\; \; \; \; \; \; \; \; \;C_8 = 0.0004,\nonumber\\
C_9&=& -0.0090,\;\; \; \; \; \; \;C_{10} = 0.0022.
\label{2k}
\vspace{2mm}
\end{eqnarray}
The combinations $a_{1}$--$a_{10}$ of the Wilson coefficients are defined as usual \cite{Ali:1998eb,Ali:1998gb,Keum:2000ms,Lu:2000hj}:
\begin{eqnarray}
a_1&=&C_2+C_1/3,\;\; \; \; \; \;a_2=C_1+C_2/3,\nonumber \\
a_i&=&C_i+C_{i\pm1}/3,\;\; \; (i=3-10),
\label{2ka}
\vspace{2mm}
\end{eqnarray}
where the upper (lower) sign applies, when $i$ is odd (even).

For the two-body decay processes of $\bar B_s^0\to M_2M_3$, we denote the emitted or annihilated meson as $M_2$ while the recoiling meson is $M_3$. The meson $M_2$ ($\rho$ or $\omega$) and the final-state meson $M_3$ ($K^{*0}$) move along the direction of $n=(1,0,{\bf{0}}_T)$ and $v=(0,1,{\bf{0}}_T)$ in the light-cone coordinates, respectively.
The decay amplitude can be expressed as the convolution of the wave functions $\phi_{B_s}$, $\phi_{M_2}$ and $\phi_{M_3}$ and the hard scattering kernel $T_H$ in the pQCD.
The pQCD factorization theorem has been developed for the two-body non-leptonic heavy meson decays, based on the formalism of Botts, Lepage, Brodsky and Sterman \cite{Chang:1996dw,Yeh:1997rq,Lepage:1980fj,Botts:1989kf}.
The basic idea of the pQCD approach is that it takes into account the
transverse momentum of the valence quarks in the hadrons which results in the Sudakov
factor in the decay amplitude. Then, the decay channels of $\bar B_s^0\to \rho^{0}(\omega)K^{*0}$ are conceptually written as the following:
\begin{eqnarray}
A(\bar B_s^0\to \rho^{0}(\omega)K^{*0})=
\int\!\! d^4k_1 d^4k_2 d^4k_3\ \mathrm{Tr} \bigl[ C(t)
\phi_{B_s}(k_1) \phi_{M2}(k_2) \phi_{M3}(k_3) T_H(k_1,k_2,k_3, t) \bigr],  \label{eq:convolution1}
\end{eqnarray}
where $k_i(i=1,2,3)$ are momentum of light quark in each meson.
$\mathrm{Tr}$ denotes the trace over Dirac structure and color indices.
$C(t)$ is the short distance Wilson coefficients at the hard scale $t$. The meson wave functions $\phi_{B_s}$ and $\phi_{M}(m=2,3)$, including all non-perturbative contribution during the hadronization of mesons, can be extracted from experimental data or other non-perturbative methods. The hard kernel $T_H(k_1,k_2,k_3, t)$ describes the four quark operator and the spectator quark connected by a hard gluon, which can be perturbatively calculated including all possible Feynman diagrams of the factorizable and non-factorizable contributions without end-point singularity.

The $\rho^{0}(\omega)$ and $K^{*0}$ mesons are treated as a light-light system. At the $B_s$ meson rest frame, they are moving very fast. We define the ratios ${r_{K^{*0}}}=\frac{M_{K^*0}}{M_{B_s}}$, ${r_\rho } = \frac{M_\rho}{M_{B_s}}$ and ${r_\omega}= \frac{M_\omega}{M_{B_s}}$. In the limit ${M_{K^{*0}} }$, ${M_\rho }$, ${M_\omega}$ $\to 0$, one can drop the terms of proportional to $r_{K^{*0}} ^2$, $r_\rho ^2$, $r_\omega ^2$ safely. The symbols $P_{B}$, $P_2$ and $P_3$ refer to the $\bar B_s^0$ meson momentum, the $\rho^{0}(\omega)$ meson momentum, and the final-state $K^{*0}$ meson momentum, respectively.
The momenta of the participating mesons in the rest frame of the $B_s$ meson can be written as:
\begin{eqnarray}
P_B&=&\frac{M_{B_s}}{\sqrt 2}(1,1,{\bf{0}}_T),\;\;
P_2=\frac{M_{B_s}}{\sqrt 2}(1,0,{\bf{0}}_T),\;\;
P_3=\frac{M_{B_s}}{\sqrt 2}(0,1, {\bf{0}}_T).
\end{eqnarray}
One can denote the light (anti-)quark momenta $k_1$, $k_2$ and $k_3$ for the initial meson $\bar B_s^0$, and the final mesons $\rho^{0}(\omega)$ and $K^{*0}$, respectively. We can choose:
\begin{eqnarray}
k_1&=&(x_1\frac{M_{B_s}}{\sqrt 2},0, {\bf k}_{1\perp}),\;\;
k_2=(x_2\frac{M_{B_s}}{\sqrt 2},0, {\bf k}_{2\perp}),\;\;
k_3=(0,x_3\frac{M_{B_s}}{\sqrt 2},{\bf k}_{3\perp}),
\end{eqnarray}
where $x_1$, $x_2$ and $x_3$ are the momentum fraction. ${\bf k}_{1\perp}$, ${\bf k}_{2\perp}$ and ${\bf k}_{3\perp}$ refer to the transverse momentum of the quark, respectively. To extract the helicity amplitudes, we parameterize the
following longitudinal polarization vectors of the $\rho^{0}(\omega)$ and $K^{*0}$ as following: 
\begin{eqnarray}
\epsilon_{2}(L)=\frac{P_{2}}{M_{\rho(\omega)}}-\frac{M_{\rho(\omega)}}{P_{2} \cdot v} v,\;\;\;\; \epsilon_{3}(L)=\frac{P_{3}}{M_{K^{*0}}}-\frac{M_{K^{*0}}}{P_{3} \cdot n} n,
\end{eqnarray}
which satisfy the orthogonality relationship of $\epsilon_{2}(L) \cdot P_{2}=\epsilon_{3}(L) \cdot P_{3}=0$, and the normalization of $\epsilon_{2}^{2}(L)=\epsilon_{3}^{2}(L)=-1$. The transverse polarization vectors can be adopted directly as
\begin{equation}
\epsilon_{2}(T)=\left(0,0,{\bf{1}}_{T}\right), \quad \epsilon_{3}(T)=\left(0,0,{\bf{1}}_{T}\right).
\end{equation}

Within the pQCD framework, both the initial and the final state meson wave functions and distribution amplitudes are important as non-perturbative input parameters. For the $B_s$ meson, the wave function of the meson  can be expressed as
\begin{equation}
 \Phi_{B_s} = \frac{i}{\sqrt{6}} (\not \! P_{B_s} +M_{B_s}) \gamma_5
\phi_{B_s} (k), \label{bmeson}
\end{equation} 
where the distribution amplitude $\phi_{B_s}$ is shown in Refs. \cite{Ali:2007ff,Li:2004ep,Wang:2014mua}:
\begin{equation}
\phi_{B_s}(x,b) = N_{B_s} x^2(1-x)^2 \exp \left[ -\frac{M_{B_s}^2\
x^2}{2 \omega_b^2} -\frac{1}{2} (\omega_b b)^2 \right].\label{waveb}
\end{equation}
The shape parameter $\omega_b$ is a free parameter and $N_{B_s}$ is a normalization factor.
Based on the studies of the light-cone sum rule, lattice QCD or be fitted to the measurements
with good precision \cite{Li:2003yj},
we take $\omega_b=0.50~\mathrm{GeV}$ for the $B_s$ meson.
The normalization factor $N_{B_s}$ depends on the values
of the shape parameter $\omega_b$ and decay constant $f_{B_s}$, which is defined through the
normalization relation $\int_0^1 {dx{\phi _{{B_s}}}\left( {x,0} \right)}  = {f_{{B_s}}}/(2\sqrt 6)$.

The distribution amplitudes of vector meson(V=$\rho$, $\omega$ or $K^{*}$), $\phi_{V}$, $\phi_{V}^T$, $\phi^t_{V}$, $\phi^s_{V}$, $\phi^v_{V}$, and $\phi^a_{V}$,
can be written in the following form \cite{Ball:1998ff,Ball:2004rg}:
\begin{eqnarray}
\phi_\rho (x)&=&\frac{3f_\rho}{\sqrt{6}} x (1-x)\left[1+0.15C_2^{3/2} (t) \right]\;,\label{phirho}\\
\phi_\omega(x)&=&\frac{3f_\omega}{\sqrt{6}} x (1-x)\left[1+0.15C_2^{3/2} (t)\right]\;,\label{phiomega}\\
\phi_{K^{*}}(x)&=&\frac{3f_{K^*}}{\sqrt{6}} x (1-x)\left[1+0.03C_1^{3/2} (t)+0.11C_2^{3/2} (t)\right]\;,\label{phik} \\
\phi_\rho^T(x)&=&\frac{3f_\rho^T}{\sqrt{6}} x (1-x)\left[1+0.14C_1^{3/2} (t) \right]\;,\label{phirho1}\\
\phi_\omega^T(x)&=&\frac{3f_\omega^T}{\sqrt{6}} x (1-x)\left[1+0.14C_1^{3/2} (t) \right]\;,\label{phirho2}\\
\phi_{K^{*}}^T(x)&=&\frac{3f_{K^*}^T}{\sqrt{6}} x (1-x)\left[1+0.04C_1^{3/2} (t)+0.10C_2^{3/2} (t) \right]\;,\label{phirho3}\\
\phi^t_V(x)&=&\frac{3f^T_V}{2\sqrt 6}t^2\;,\label{phitv}\\
\phi^s_V(x)&=&\frac{3f_V^T}{2\sqrt 6} (-t)\;,\label{phisv}\\
\phi_V^v(x)&=&\frac{3f_V}{8\sqrt6}(1+t^2)\;,\label{phivv}\\
 \phi_V^a(x)&=&\frac{3f_V}{4\sqrt6}(-t)\;,\label{phiav}
\end{eqnarray}
where $t=2x-1$. Here $f_{V}^{(T)}$ is the decay constant of
the vector meson with longitudinal(transverse) polarization. The Gegenbauer polynomials $C^{\nu}_{n}(t)$ can be
defined as \cite{Fan:2012kn,Huang:2005if}:
\begin{eqnarray}
C^{3/2}_{1}(t)&=&3t  \\
C_2^{3/2} (t)&=&\frac{3}{2}(5t^2-1).
\end{eqnarray}

\section{\label{sec:cpv1}$CP$ violation in $\bar{B}_s^0\to\rho^{0}(\omega )K^{*0}\to {\pi^+}{\pi^-}K^{*0}$ decay process}
\subsection{\label{subsec:form}Formalism}

The decay width $\Gamma$ for the processes of $\bar{B}_s^0\to\rho^{0}(\omega ) K^{*0}$ is given by 
\begin{equation}
\Gamma  = \frac{{{P_c}}}{{8\pi M_{{B_s}}^2}}\sum\limits_{\sigma=L,T}{{A^{(\sigma) \dagger }}{A^{(\sigma )}}}, \label{dr1}
\end{equation}
where $P_c$ is the absolute value of the three-momentum of the final state mesons. The decay amplitude $A^{(\sigma)}$ which is decided by QCD dynamics will be calculated later in pQCD factorization approach. The superscript $\sigma$ denotes the
helicity states of the two vector mesons with the longitudinal (transverse) components L(T). The amplitude
$A^{(\sigma)}$ for the decays $B_s(P_{B_s})\to V_{\rho(\omega)}(P_2,\epsilon^*_{2\mu})+V_{K^{*0}}(P_3,\epsilon^*_{3\mu})$ can be decomposed as follows \cite{Zhu:2005rx,Huang:2005if,Lu:2005be,Li:2004ti}:
\begin{eqnarray}
A^{(\sigma)}=M^2_{B_{s}}A_{L}+M^2_{B_{s}}A_{N}
\epsilon^{*}_{2}(\sigma=T)\cdot\epsilon^{*}_{3}(\sigma=T) +i
A_{T}\epsilon^{\alpha \beta\gamma \rho}
\epsilon^{*}_{2\alpha}(\sigma)\epsilon^{*}_{3\beta}(\sigma)
P_{2\gamma }P_{3\rho }\;,
\end{eqnarray}
where $\epsilon^*$ is the polarization vector of the vector meson. The amplitude $A_{i}$
($i$ refer to the three kinds of polarizations, longitudinal (L), normal (N) and transverse (T)) can be written as
\begin{eqnarray}
M^2_{B_{s}}A_L &=& a \,\, \epsilon_2^{*}(L) \cdot \epsilon_3^{*}(L) +{\frac{b}{M_2 M_{3}}} \epsilon_{2}^{*}(L) \cdot P_3 \,\, \epsilon_{3}^{*}(L) \cdot P_2\;, \nonumber \\
M^2_{B_{s}}A_N &=& a \;, \label{id-rel}  \nonumber \\
A_T &=& {\frac{c}{M_2 M_{3}}}\;,
\end{eqnarray}
where $a$, $b$ and $c$ are the Lorentz-invariant amplitudes. $M_2$ and $M_3$ are the masses of the vector mesons $\rho^{0}(\omega)$ and $K^{*0}$, respectively.

The longitudinal $H_{0}$ and transverse $H_{\pm}$ of helicity amplitudes can be expressed
\begin{eqnarray}
H_{0}&=&M^2_{B_{s}}A_{L}, \nonumber\\
H_{\pm}&=&M^2_{B_{s}}A_{N}\mp M_{2}M_{3} \sqrt{\kappa^2-1}A_{T},
\label{dr1bbbbbbbbb}
\end{eqnarray}
where
 $H_{0}$ and $H_{\pm}$ are the penguin-level and tree-level helicity amplitudes of the decay process $\bar B_s^0 \to \rho (\omega ) K^{*0}  \to {\pi ^ + }{\pi ^ - }K^{*0} $ from the three kinds of polarizations, respectively. The helicity summation satisfy the relation 
\begin{equation}
\sum\limits_{\sigma=L,R}{{A^{(\sigma) \dagger }}{A^{(\sigma )}}}=|H_{0}|^{2}+|H_{+}|^{2}+|H_{-}|^{2}. \label{dr1aaa}
\end{equation}

In the vector meson dominance model \cite{Nambu:1997vw,Sakurai:1969zz}, the photon propagator is dressed by coupling to vector mesons. 
Based on the same mechanism, $\rho-\omega$ mixing was proposed and later gradually applied to $B$ meson physics \cite{Gardner:1997qk,OConnell:1995nse,Lu:2010vb,Gardner:1997yx}.
 According to the effective Hamiltonian, the amplitude $A$ ($\bar{A}$) for the three-body decay process 
$\bar B_s^0 \to  {\pi ^ + }{\pi ^ - } K^{*0}$
($B_s^0 \to {\pi ^ + }{\pi ^ - } \bar K^{*0}$) can be written as \cite{Gardner:1997yx}:
\begin{eqnarray}
A&=&\big<\pi^+\pi^{-}K^{*0}|H^T|\bar{B}_{s}^{0}\big>+\big<\pi^+\pi^{-}K^{*0}|H^P|\bar{B}_{s}^{0}\big>,\label{A}\\
\bar{A}&=&\big<\pi^+\pi^{-}\bar K^{*0}|H^T|{B}_{s}^{0}\big>+\big<\pi^+\pi^-\bar  K^{*0}|H^P|{B}_{s}^{0}\big>,\label{Abar}
\end{eqnarray}
where $H^T$ and $H^P$ are the Hamiltonian for the tree and penguin operators, respectively.

The relative magnitude and phases between the tree
and penguin operator contribution are defined as follows:
\begin{eqnarray}
A&=&\big<\pi^+\pi^{-} K^{*0}|H^T|\bar{B}_{s}^{0}\big>[1+re^{i(\delta+\phi)}],\label{A'}\\
\bar{A}&=&\big<\pi^+\pi^-\bar K^{*0}|H^T|{B}_{s}^{0}\big>[1+re^{i(\delta-\phi)}],
\label{A'bar}
\end{eqnarray}
where $\delta$ and $\phi$ are strong and weak phases differences, respectively.
The weak phase difference $\phi$ can be expressed as a combination of the CKM matrix elements, and it is
$\phi=\arg[(V_{tb}V_{td}^{*})/(V_{ub}V_{ud}^{*})]$ for the $b \to d$ transition. The parameter $r$ is the absolute value of the ratio of tree and penguin amplitudes:
\begin{eqnarray}
r\equiv\Bigg|\frac{\big<\pi^+\pi^{-}K^{*0}|H^P|\bar{B}_{s}^{0}\big>}{\big<\pi^+\pi^{-}K^{*0}|H^T|\bar{B}_{s}^{0}\big>}\Bigg|
\label{r}.
\end{eqnarray}
The parameter of $CP$ violating asymmetry, $A_{CP}$, can be written as
\begin{eqnarray}
A_{CP}=\frac{|A|^{2}-|\bar{A}|^{2}}{|A|^{2}+|\bar{A}|^{2}} =\frac{-2(T_{0}^2r_{0}\sin\delta_0+T_{+}^2r_{+}\sin\delta_{+}+T_{-}^2r_{-}\sin\delta_-)\sin\phi}  {\sum_{i=0+-}T_{i}^2(1+r_{i}^2+2r_{i}\cos\delta_i\cos\phi)},
\label{eq:CP-tuidao}
\end{eqnarray}
where $T_{i}(i=0,+,-)$ represent the tree-level helicity amplitudes of the decay process $\bar B_s^0 \to  {\pi ^ + }{\pi ^ - } K^{*0}$ from $H_{0}$, $H_{+}$ and $H_{-}$ of the Eq. (\ref{dr1bbbbbbbbb}), respectively. $r_{j}(j=0,+,-)$ refer to the absolute value of the ratio of tree and penguin amplitude for the three kinds of polarizations, respectively.
$\delta_{k}(k=0,+,-)$ are the relative strong phases between the tree
and penguin operator contributions from three kinds of helicity amplitudes, respectively.
We can see explicitly from Eq. (\ref{eq:CP-tuidao}) that both weak and strong phase
differences are needed to produce $CP$ violation. In order to obtain a large signal for direct $CP$ violation, we intend to apply the $\rho-\omega$ mixing mechanism, which leads to large strong phase differences in hadron decays. 

\begin{figure}[tbh]
	\centering
	\includegraphics[width=0.3\textwidth]{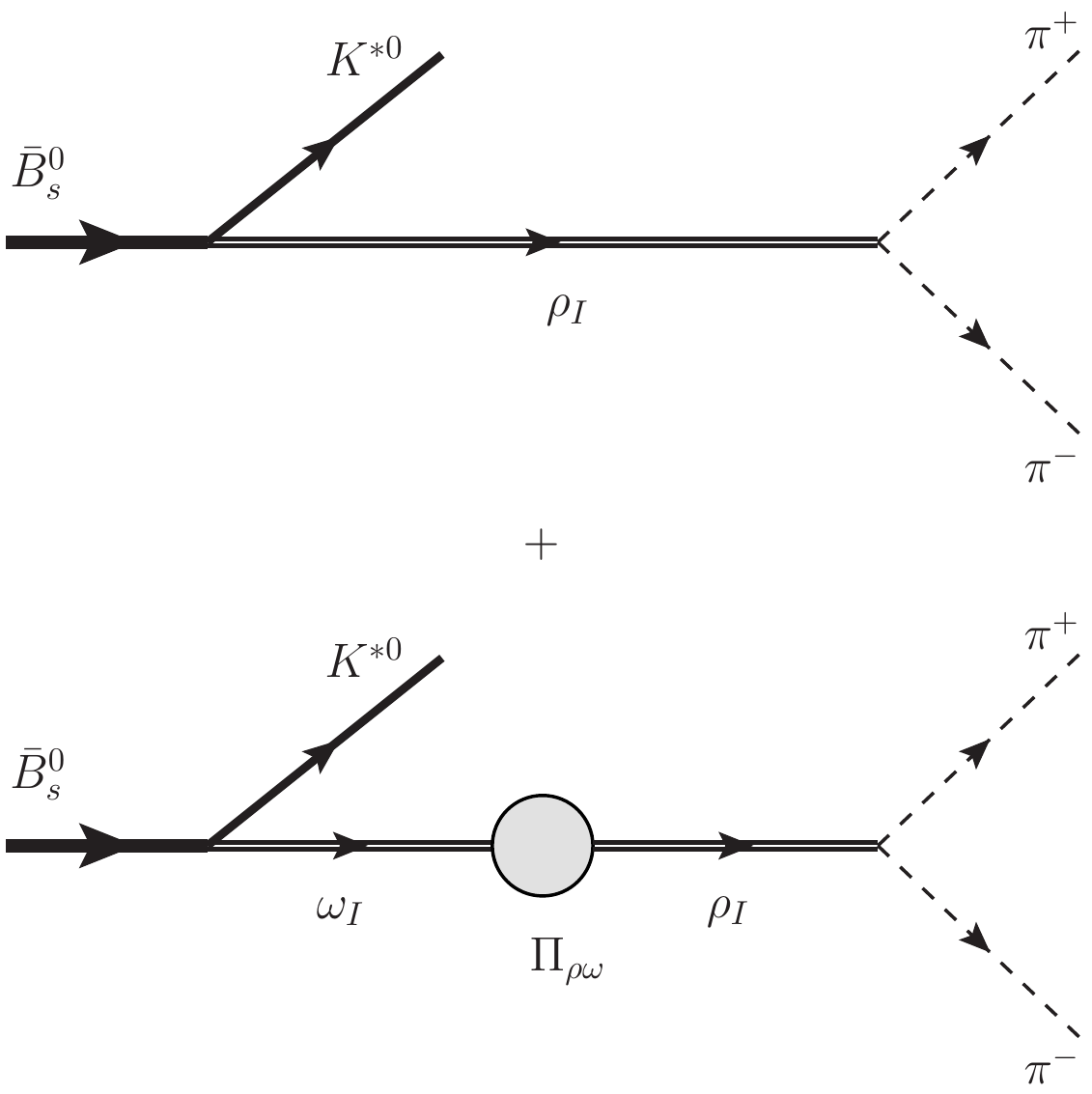}
	\caption{\label{rho-omega} The diagram for the $\bar{B}^0_{s}\to {\pi^+}{\pi^-}K^{*0}$ decay with the $\rho-\omega$ mixing mechanism  for the first order of isospin violation in the isospin representation}
\end{figure}
With the $\rho-\omega$ mixing mechanism, the process of the $\bar{B}^0_{s}\to\rho(\omega) K^{*0} \to \pi^+\pi^-K^{*0}$ decay is shown in Fig.\ref{rho-omega}. In the isospin representation, the decay amplitude $M_{\bar{B}_s^0}\to \rho(\omega) \to \pi^+\pi^-$ in Fig.1 can be written as \cite{Lu:2013jma,Guo:1999ip,Guo:2008zzh,OConnell:1995nse}
\begin{eqnarray}
M_{\bar{B}_s^0}\to \pi^+\pi^-=M_{\rho_I \to \pi\pi}\frac{1}{s_\rho}M_{\bar{B}_s^0 \to \rho_I}+M_{\rho_I \to \pi\pi}\frac{1}{s_\rho}\Pi_{\rho\omega}\frac{1}{s_\omega}M_{\bar{B}_s^0 \to \omega_I}.
\end{eqnarray}
Introducing the $\epsilon=\frac{\Pi_{\rho\omega}}{s_\rho-s_\omega}$ \cite{Lu:2013jma,Guo:1999ip,Guo:2008zzh,OConnell:1995nse}, We have identified the physical amplitudes as
\begin{eqnarray} 
M_{\rho \to \pi\pi}&=&M_{\rho_I \to \pi\pi}, \\
M_{\omega \to \pi\pi}&=&\epsilon M_{\rho_I \to \pi\pi}, \\
M_{\bar{B}_s^0 \to \rho}&=&M_{\bar{B}_s^0 \to \rho_I}-\epsilon M_{\bar{B}_s^0 \to \omega_I}, \\
M_{\bar{B}_s^0 \to \omega}&=&M_{\bar{B}_s^0 \to \omega_I}-\epsilon M_{\bar{B}_s^0 \to \rho_I}. 
\end{eqnarray}
From the physical representation, we can obtain the decay amplitude
\begin{eqnarray}
M_{\bar{B}_s^0}\to \pi^+\pi^-
=M_{\rho \to \pi\pi}\frac{1}{s_\rho}M_{\bar{B}_s^0 \to \rho}+M_{\rho \to \pi\pi}\frac{\Pi_{\rho\omega} s_\rho}{s_\rho -s_\omega }\frac{1}{s_\rho s_\omega}M_{\bar{B}_s^0 \to \omega_I}
\end{eqnarray}
where $\mathcal{O} (\epsilon^{2})$ corrections is neglected, and $M_{\rho \to \pi\pi}=g_\rho$, $M_{\bar{B}_s^0 \to \rho}$=$t_\rho^i$ or $p_\rho^i$ and $M_{\bar{B}_s^0 \to \omega}$=$t_\omega^i$ or $p_\omega^i$ are used. So, we can get $\widetilde{\Pi}_{\rho\omega}$=$\frac{\Pi_{\rho\omega} s_\rho}{s_\rho -s_\omega }$. $\widetilde{\Pi}_{\rho\omega}$ is the effective $\rho-\omega$ mixing amplitude which also effectively includes the direct coupling $\omega\rightarrow\pi^+\pi^-$. 
At the first order of isospin violation, we have the following tree
and penguin amplitudes when the invariant mass of $\pi^+\pi^-$ pair is near the $\omega$ resonance mass \cite{Gardner:1997yx,Guo:1998eg}:
\begin{eqnarray}
\big<\pi^+\pi^-K^{*0}|H^T|\bar{B}^0_{s}\big>=\frac{g_{\rho}}{s_{\rho}s_{\omega}}\widetilde{\Pi}_{\rho\omega}t_{\omega}^{i}+\frac{g_{\rho}}{s_{\rho}}t_{\rho}^{i},
\label{Htr}\\
\big<\pi^+\pi^-K^{*0}|H^P|\bar{B}^0_{s}\big>=\frac{g_{\rho}}{s_{\rho}s_{\omega}}\widetilde{\Pi}_{\rho\omega}p_{\omega}^{i}+\frac{g_{\rho}}{s_{\rho}}p_{\rho}^{i},
\label{Hpe}
\end{eqnarray}
where $t_{\rho}^{i}(p_{\rho}^{i})$ and $t_{\omega}^{i}(p_{\omega}^{i})$ are the tree (penguin)-level helicity amplitudes
for $\bar{B}_{s}^0\rightarrow\rho^0 K^{*0}$ and
$\bar{B}_{s}^0\rightarrow\omega K^{*0}$, respectively. The amplitudes $t_{\rho}^{i}$,  $t_{\omega}^{i}$, $p_{\rho}^{i}$ and $p_{\omega}^{i}$ can be
found in Sec. \ref{cal}. 
$g_{\rho}$ is the coupling constant for the decay process $\rho^0\rightarrow\pi^+\pi^-$. $s_{V}$, $m_{V}$ and $\Gamma_V$($V$=$\rho$ or
$\omega$) is the inverse propagator, mass and decay width of the vector meson $V$, respectively. $s_V$ can be expressed as
\begin{eqnarray}
s_V=s-m_V^2+{\rm{i}}m_V\Gamma_V,
\end{eqnarray}
with $\sqrt{s}$ being the invariant masses of the $\pi^+\pi^-$ pairs. The  $\rho- \omega$ mixing parameter $\widetilde{\Pi}_{\rho\omega}(s)={\rm{Re}}\widetilde{\Pi}_{\rho\omega}(m_{\omega}^2)+{\rm{Im}} \widetilde{\Pi}_{\rho\omega}(m_{\omega}^2)$ are \cite{Lu:2018fqe}
\begin{eqnarray}
{\rm{Re}} \widetilde{\Pi}_{\rho\omega}(m_{\omega}^2)&=&-4760\pm{440}\,
\rm{MeV}^2,\nonumber\\ {\rm{Im}} \widetilde{\Pi}_{\rho\omega}(m_{\omega}^2)&=&-6180\pm{3300}\,
\textrm{MeV}^2. \label{rhoomegamixing}
\end{eqnarray}
From Eqs. (\ref{A}), (\ref{A'}), (\ref{Htr}) and (\ref{Hpe}) one has
\begin{eqnarray}
re^{i\delta_{i}}e^{i\phi}=\frac{\widetilde{\Pi}_{\rho\omega}p_{\omega}^{i}+s_{\omega}p_{\rho}^{i}}{\widetilde{\Pi}_{\rho\omega}t_{\omega}^{i}+s_{\omega}t_{\rho}^{i}},
\label{rdtdirive}
\end{eqnarray}
Defining \cite{Enomoto:1996cv,Enomoto:1997bq}
\begin{eqnarray}
\frac{p_{\omega}^{i}}{t_{\rho}^{i}}\equiv r^\prime
e^{i(\delta^{i}_q+\phi)},\quad\frac{t_{\omega}^{i}}{t_{\rho}^{i}}\equiv
\alpha
e^{i\delta^{i}_\alpha},\quad\frac{p_{\rho}^{i}}{p_{\omega}^{i}}\equiv
\beta e^{i\delta^{i}_\beta}, \label{def}
\end{eqnarray}
where $\delta^{i}_\alpha$, $\delta^{i}_\beta$ and $\delta^{i}_q$ are strong
phases of the decay process $\bar B_s^0 \to \rho^{0} (\omega ) K^{*0}  \to {\pi ^ + }{\pi ^ - }K^{*0} $ from the three kinds of polarizations, respectively. One finds the following expression from Eqs. (\ref{rdtdirive}) and (\ref{def}):
\begin{eqnarray}
re^{i\delta_{i}}=r^\prime
 e^{i\delta^{i}_q}\frac{\widetilde{\Pi}_{\rho\omega}+\beta
e^{i\delta^{i}_\beta}s_{\omega}}{\widetilde{\Pi}_{\rho\omega}\alpha
e^{i\delta^{i}_\alpha}+s_{\omega}}. \label{rdt}
\end{eqnarray}
$\alpha e^{i\delta^{i}_\alpha}$, $\beta e^{i\delta^{i}_\beta}$, and $r^\prime e^{i \delta^{i}_q}$ will be calculated in the perturbative QCD approach. In order to obtain the $CP$ violating asymmetry in Eq.
(\ref{eq:CP-tuidao}), $A_{CP}$, sin$\phi$ and cos$\phi$ are needed. $\phi$ is
determined by the CKM matrix elements. In the Wolfenstein
parametrization \cite{Wolfenstein:1964ks}, the weak phase $\phi$ comes from $[{V_{tb}}V_{td}^*/{V_{ub}}V_{ud}^*]$.
One has
\begin{eqnarray}
{\rm sin}\phi &=&\frac{\eta }{{\sqrt {{{\left[ {\rho \left( {1 - \rho } \right) - {\eta ^2}} \right]}^2} + {\eta ^2}} }},  \label{sin1} \\
{\rm cos}\phi &=&\frac{{\rho \left( {1 - \rho } \right) - {\eta ^2}}}{{\sqrt {{{\left[ {\rho \left( {1 - \rho } \right) - {\eta ^2}} \right]}^2} + {\eta ^2}} }},
\label{3l1}
\vspace{2mm}
\end{eqnarray}
where the same result has been used for $b \to d$ transition from Ref. \cite{Ajaltouni:2003yt,Leitner:2002xh}.

\subsection{\label{cal}Calculation details}

We can decompose the decay amplitudes for the decay processes $\bar{B}_{s}^0\rightarrow \rho^{0}(\omega)K^{*0}$
in terms of tree and penguin contributions depending on the CKM matrix elements of $V_{ub}V^{*}_{ud}$ and $V_{tb}V^{*}_{td}$.
From Eqs. (\ref{eq:CP-tuidao}), (\ref{rdtdirive}) and (\ref{def}), in leading order to obtain the formulas of the $CP$ violation, we need calculate the amplitudes $t_{\rho}$, $p_{\rho}$, $t_{\omega}$ and $p_{\omega}$ in perturbative QCD approach. The relevant function can be found in the Appendix from the perturbative QCD approach.

\begin{figure}[tbh]
	\centering
	\includegraphics[width=0.18\textwidth]{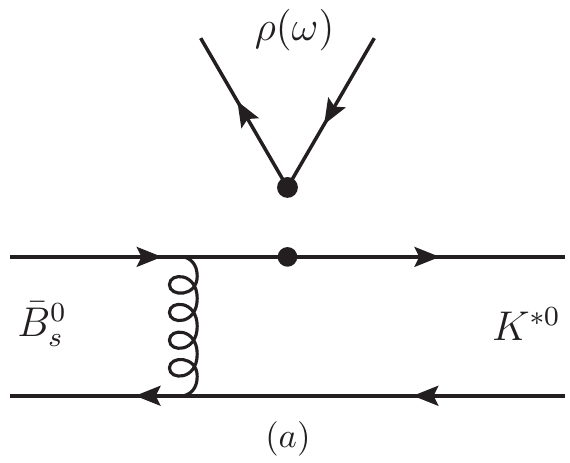}
	\hspace{0.15in}
	\includegraphics[width=0.18\textwidth]{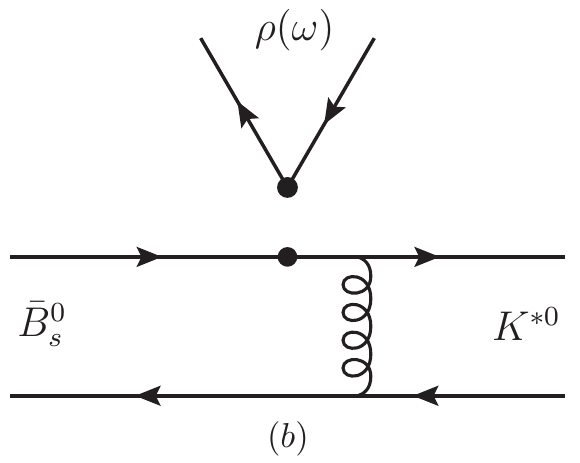}
	\hspace{0.15in}	
	\includegraphics[width=0.18\textwidth]{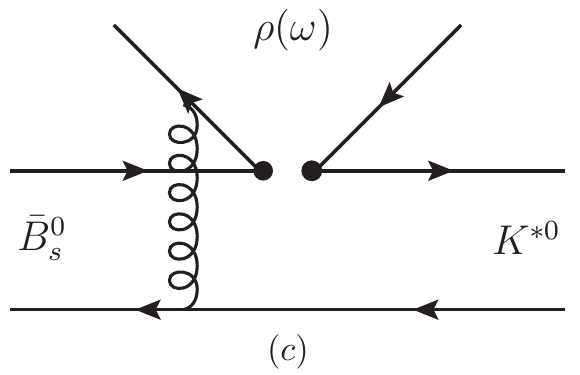}
	\hspace{0.15in}
	\includegraphics[width=0.18\textwidth]{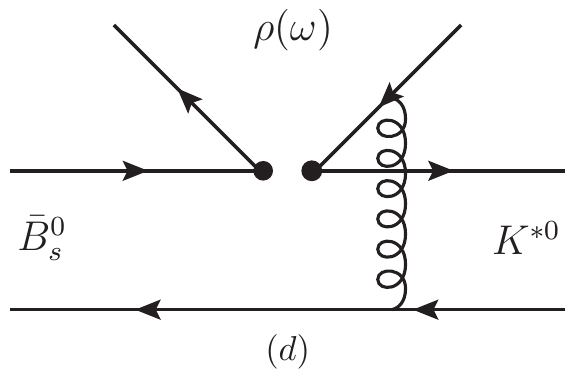}	\\  \vspace{1em}
	\includegraphics[width=0.18\textwidth]{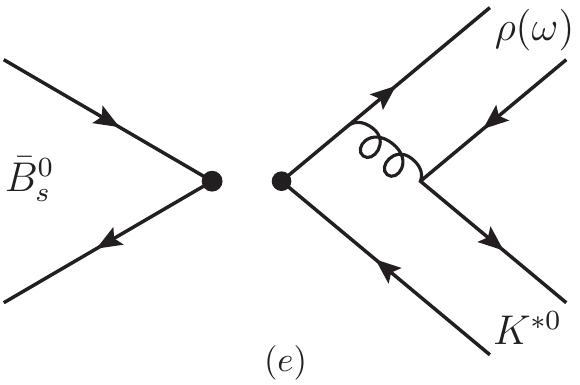}
	\hspace{0.15in}
	\includegraphics[width=0.18\textwidth]{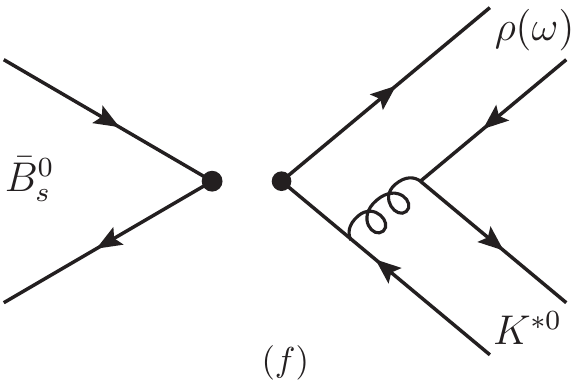}
	\hspace{0.15in}	
	\includegraphics[width=0.18\textwidth]{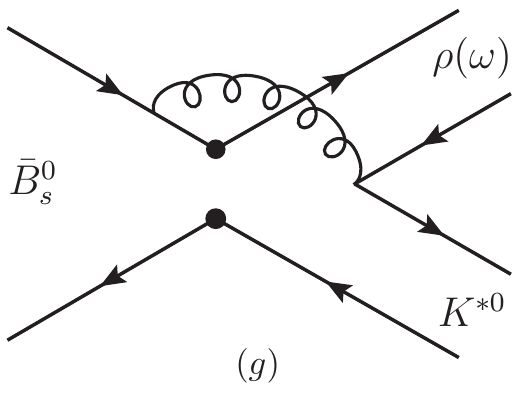}
	\hspace{0.15in}
	\includegraphics[width=0.18\textwidth]{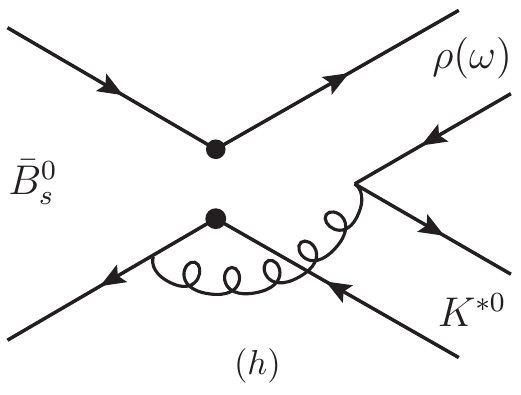}
	\caption{\label{emission plot} Leading order Feynman diagrams for $\bar{B}_s^0 \to \rho^0(\omega)K^{*0}$} 
\end{figure}
In the pQCD, there are eight types of the leading order Feynman diagrams contributing to $\bar{B}_{s}^0\rightarrow \rho^{0}(\omega)K^{*0}$ decays, which are shown in Fig.\ref{emission plot}. The first row is for the emission-type diagrams, where the first two diagrams in Fig.\ref{emission plot} (a)(b) are called factorizable emission diagrams and the last two diagrams in Fig.\ref{emission plot} (c)(d) are called non-factorizable emission diagrams \cite{Zhu:2005rx,Chen:2002pz}. The second row is for the annihilation-type diagrams, where the first two diagrams in Fig.\ref{emission plot} (e)(f) are called factorizable annihilation diagrams and the last two diagrams in Fig.\ref{emission plot} (g)(h) are called non-factorizable annihilation diagrams \cite{Shen:2006ms,Wang:2014mua}.
The relevant decay amplitudes can be easily obtained by these hard gluon exchange diagrams and the Lorenz structures of the mesons wave functions.
Through calculating these diagrams, the formulas of $\bar{B}_{s}^0 \to \rho K^{*0}$ or $\bar{B}_{s}^0 \to \omega K^{*0}$ are similar to those of $B\to \phi K^*$ and $B_s \to K^{*-}K^{*+}$ \cite{Chen:2002pz,Zou:2015iwa}.
We just need to replace some corresponding Wilson coefficients, wave functions and corresponding parameters.

With the Hamiltonian equation (\ref{2a}), depending on CKM matrix elements of $V_{ub}V^{*}_{ud}$ and  $V_{tb}V^{*}_{td}$,
the tree dominant decay amplitudes $A^{(i)}$ for $\bar B_{s}^0\rightarrow \rho K^{*0}$ in pQCD can be
written as
\begin{eqnarray}
\sqrt{2}A^{(i)}(\bar B_{s}^0\to\rho^{0}K^{*0})&=&V_{ub}V_{ud}^{*}T_{\rho}^{i}-V_{tb}V_{td}^{*}P_{\rho}^{i} , \label{BcDrho1}
\end{eqnarray}
where the superscript $i$ denote different helicity
amplitudes $L,N$ and $T$. The longitudinal $t_{\rho(\omega) }^0$, transverse $t_{\rho(\omega) }^{\pm}$ of helicity amplitudes satisfy relationship from Eq. (\ref{dr1bbbbbbbbb}). The amplitudes of the tree and penguin diagrams can be written as 
 $T_\rho ^i = t_\rho ^i/{V_{ub}}V_{ud}^*$ and $P_\rho ^i = p_\rho ^i/{V_{tb}}V_{td}^*$, respectively. 
The formula for the tree level amplitude is
\begin{eqnarray}
T_{\rho}^{i}=\frac{G_F}{\sqrt{2}}\Big \{ f_{\rho} F_{
B_s\to K^{*}}^{LL,i} \left[a_{2}\right]+ M_{ B_s\to
K^{*}}^{LL,i}\left[C_{2}\right]\Big \},  \label{trho1}
\end{eqnarray}
where $f_{\rho}$ refers to the decay constant of $\rho$ meson. The penguin level amplitude are expressed in the following
\begin{eqnarray}   
P_{\rho}^{i}&=&-\frac{G_F}{\sqrt{2}}\Big \{f_{\rho} F_{B_s\to K^*}^{LL,i} \left[
-a_{4}+\frac{3}{2}a_7+\frac{3}{2}a_9+\frac{1}{2}a_{10}\right] -  M_{B_s\to K^*}^{LR,i} \left[-C_{5}+\frac{1}{2}C_{7}\right] \nonumber\\
&& +  M_{B_s\to K^*}^{LL,i} \left[-C_{3}+\frac{1}{2}C_{9}+\frac{3}{2}C_{10}\right] -M_{B_s\to K^*}^{SP,i} \left[\frac{3}{2}C_{8}\right]+  f_{B_s}  F_{ann}^{LL,i}\left[-a_{4}+ \frac{1}{2}a_{10}\right] \nonumber\\
&& 
-  f_{B_s} F_{ann}^{SP,i}\left[-a_{6}+\frac{1}{2}a_{8}\right]
 +  M_{ann}^{LL,i}\left[-C_{3}+\frac{1}{2}C_{9}\right]
-  M_{ann}^{LR,i}\left[-C_{5}+\frac{1}{2}C_{7}\right]\bigg \}.  \label{Prho1}
\end{eqnarray}
The tree dominant decay amplitude for $\bar B_{s}^0\to \omega K^{*0}$ can be written as
\begin{eqnarray}
\sqrt 2A^i(\bar B_{s}^0\to \omega K^{*0})= V_{ub}V_{ud}^{*} T_{\omega}^{i} -  V_{tb}V_{td}^{*} P_{\omega}^{i}, \label{BcDomega1}
\end{eqnarray}
where $T_\omega ^i = t_\omega ^i/{V_{ub}}V_{us}^*$ and $P_\omega ^i = p_\omega ^i/{V_{tb}}V_{ts}^*$ which refer to the tree and penguin amplitude, respectively.
We can give the tree level contribution in the following
\begin{eqnarray}
T_{\omega}^{i}=\frac{G_F}{\sqrt{2}}\Big\{f_\omega F_{B_s\to
K^*}^{LL,i}\left[a_{2}\right] + M_{B_s\to
K^*}^{LL,i}\left[C_{2}\right]\Big \}, \label{tomega1}
\end{eqnarray}
where $f_{\omega}$ refers to the decay constant of $\omega$ meson.
The penguin level contribution are given as following
\begin{eqnarray}
P_{\omega}^{i}&=& \frac{G_F}{\sqrt{2}} \bigg\{ f_{\omega} F_{B_s\to K^*}^{LL,i}  \left[
2a_{3}+a_{4}+2a_{5}
+\frac{1}{2}a_{7}+\frac{1}{2}a_{9}-\frac{1}{2}a_{10}\right]- M_{B_s\to K^*}^{LR,i}\left[C_{5}-\frac{1}{2}C_{7}\right] \nonumber\\
&&   + M_{B_s\to K^*}^{LL,i}\left[C_{3}+2C_{4}-\frac{1}{2}C_9 +\frac{1}{2}C_{10}\right]   
- M_{B_s\to K^*}^{SP,i}\left[2C_{6}+\frac{1}{2}C_{8}\right] + f_{B_s}  F_{ann}^{LL,i}\left[a_{4} -\frac{1}{2}a_{10}\right] \nonumber\\
&& -f_{B_s} F_{ann}^{SP,i}\left[a_{6} -\frac{1}{2}a_{8}\right]  + M_{ann}^{LL,i}\left[C_{3}-\frac{1}{2}C_{9}\right]- M_{ann}^{LR,i}\left[C_{5}-\frac{1}{2}C_{7}\right] \bigg\} \label{pomega1}
\end{eqnarray}
Based on the definition of Eq. (\ref{def}), we can get
\begin{eqnarray}
\alpha e^{i\delta^{i}_\alpha}&=&\frac{t_{\omega}^{i}}{t_{\rho}^{i}}, \label{eq:afaform} \\
\beta e^{i\delta^{i}_\beta}&=&\frac{p_{\rho}^{i}}{p_{\omega}^{i}}, \label{eq:btaform}\\
r^\prime e^{i\delta^{i}_q}&=&\frac{P^{i}_{\omega}}{T^{i}_{\rho}}
\times\bigg|\frac{V_{tb}V_{td}^*}{V_{ub}V_{ud}^*}\bigg|,  \label{eq:delform}
\end{eqnarray}
where
\begin{eqnarray}
\left|\frac{V_{tb}V^{*}_{td}}{V_{ub}V^{*}_{ud}}\right|=\frac{{\sqrt {{{\left[ {\rho \left( {1 - \rho } \right) - {\eta ^2}} \right]}^2} + {\eta ^2}} }}{{\left( {1 - {\lambda ^2}/2} \right)\left( {{\rho ^2} + {\eta ^2}} \right)}}.
\label{3p}
\vspace{2mm}
\end{eqnarray}
From above equations, the new strong phases ${\delta^{i} _\alpha }$, $\delta^{i}_\beta$ and $\delta^{i}_q$ are obtained from tree and penguin diagram contributions by the $\rho-\omega$ interference. Substituting Eqs. (\ref{eq:afaform}), (\ref{eq:btaform}) and (\ref{eq:delform}) into (\ref{rdt}), we can obtain total
strong phase $\delta_{i}$ in the framework of pQCD. Then in combination with Eqs. (\ref{sin1}) and (\ref{3l1}) the $CP$ violating asymmetry can be obtained.

\section{\label{Br}BRANCHING RATIO OF $\bar{B}^0_{s}\rightarrow \rho^{0}(\omega)K^{*0}$}

Based on the relationship of Eqs. (\ref{dr1}) and (\ref{dr1aaa}), we can calculate the decay rates for the processes of $\bar{B}_s^0\to\rho^{0}(\omega ) K^{*0}$ by using the following expression: 
\begin{equation}
\Gamma  = \frac{{{P_c}}}{{8\pi M_{{B_s}}^2}}(|H_{0}|^{2}+|H_{+}|^{2}+|H_{-}|^{2}), \label{gamma2}
\end{equation}
where
\begin{eqnarray}
P_c=\frac{\sqrt{[M_{B_s}-(M_{\rho/\omega}+M_{K^{*0}})^2][M_{B_s}-(M_{\rho/\omega}-M_{K^{*0}})^2]}}{2M_{B_s}}
\end{eqnarray}
is the c.m. momentum of the product particle and $H_i(i=0,+,-)$ are helicity amplitudes.

In this case we take into account the $\rho-\omega$ mixing contribution to the branching ratio, since we are working to the first order of isospin violation. The derivation is straightforward and we can explicitly express the branching ratio for the processes $\bar{B}_s^0\to\rho^{0}(\omega) K^{*0}$ \cite{Leitner:2002xh,Lu:2013xea}:  
\begin{eqnarray}
BR(\bar{B}_s^0\to\rho^{0}(\omega) K^{*0})=\frac{\tau_{B^0_s} P_c}{{8\pi M_{{B_s}}^2}}(|H_{\rho\omega0}|^{2}+|H_{\rho\omega+}|^{2}+|H_{\rho\omega-}|^{2}),
\end{eqnarray}
where
$\tau_{B^0_s}$ is the lifetime of the $B_s$ meson and 
\begin{eqnarray}
H_{\rho\omega i(i=0,+,-)}=(|V_{ub}V_{ud}^{*}|T_{\rho}^{i}-|V_{tb}V_{td}^{*}|P_{\rho}^{i})+ (|V_{ub}V_{ud}^{*}| T_{\omega}^{i} -  |V_{tb}V_{td}^{*}| P_{\omega}^{i})\frac{\widetilde{\Pi}_{\rho\omega}}{(s_\rho-M_\omega^2)+{\rm{i}}M_\omega\Gamma_\omega}
\end{eqnarray}
take into account the helicity amplitudes of the $\rho$ meson and $\omega$ meson contribution involved in the tree and penguin diagrams.

\section{\label{int}Input parameters}

The CKM matrix, which elements are determined from experiments, can be expressed in terms of the Wolfenstein parameters $A$, $\rho$, $\lambda$ and $\eta$ \cite{Wolfenstein:1983yz,Wolfenstein:1964ks}:
\begin{equation}
\left(
\begin{array}{ccc}
  1-\frac{1}{2}\lambda^2   & \lambda                  &A\lambda^3(\rho-\mathrm{i}\eta) \\
  -\lambda                 & 1-\frac{1}{2}\lambda^2   &A\lambda^2 \\
  A\lambda^3(1-\rho-\mathrm{i}\eta) & -A\lambda^2              &1\\
\end{array}
\right),\label{ckm}
\end{equation}
where $\mathcal{O} (\lambda^{4})$ corrections are neglected. The latest values for the parameters in the CKM matrix are \cite{Zyla:2020zbs}:
\begin{eqnarray}
&& \lambda=0.22650\pm 0.00048,\quad A=0.790_{-0.012}^{+0.017},\nonumber \\
&& \bar{\rho}=0.141_{-0.017}^{+0.016},\quad
\bar{\eta}=0.357\pm 0.011,\label{eq: rhobarvalue}
\end{eqnarray}
where
\begin{eqnarray}
 \bar{\rho}=\rho(1-\frac{\lambda^2}{2}),\quad
\bar{\eta}=\eta(1-\frac{\lambda^2}{2}).\label{eq: rho rhobar
relation}
\end{eqnarray}
From Eqs. (\ref{eq: rhobarvalue}) and (\ref{eq: rho rhobar relation})
we have
\begin{eqnarray}
0.127<\rho<0.161,\quad  0.355<\eta<0.378.\label{eq: rho value}
\end{eqnarray}
%
\begin{table*}[t]
\setlength{\abovecaptionskip}{0pt}
\setlength{\belowcaptionskip}{8pt}
\begin{center}
\renewcommand\arraystretch{1.5}
\tabcolsep 0.25in
\caption{Input parameters} \label{table2}
\begin{tabular}{lll}
\hline \hline
Parameters&Input data & References  \\ \hline
Fermi constant (in $\rm{GeV}^{-2}$)&$G_F=1.16638\times10^{-5}.$& \cite{Zyla:2020zbs}\\
                        &$M_{B^0_s}=5366.88,~\tau_{B^0_s}=1.515\times10^{-12}s,$& \\
                        &$M_{\rho^0(770)}=775.26, ~\Gamma_{\rho^0(770)}=149.1,$&\\
Masses and decay widths (in MeV)  &$M_{\omega(782)}=782.65, ~\Gamma_{\omega(782)}=8.49,$& \cite{Zyla:2020zbs}\\
                        &$M_\pi=139.57,~M_{K*}=895.55.$&\\
                       &$f_\rho=215.6\pm5.9,~f_\rho^T=165\pm9,$&\\
Decay constants (in MeV)   &$f_\omega=196.5\pm4.8,~f_\omega^T=145\pm10,$& \cite{Straub:2015ica,Liu:2016rqu,Ball:2004rg}\\
               &$f_{K^{*}}=217\pm5,~f_{K^*}^T=185\pm10.$& \\ \hline \hline
\end{tabular}
\end{center}
\end{table*}
\section{\label{num}The numerical results of $CP$ violation and Branching ratio}

\subsection{\label{num:CP}$CP$ violation via $\rho-\omega$ mixing in $\bar{B}^0_{s}\rightarrow \rho^0(\omega) K^{*0}\rightarrow\pi^+\pi^- K^{*0}$}

\begin{figure}[h]
  \centering
\includegraphics[width=0.5\textwidth]{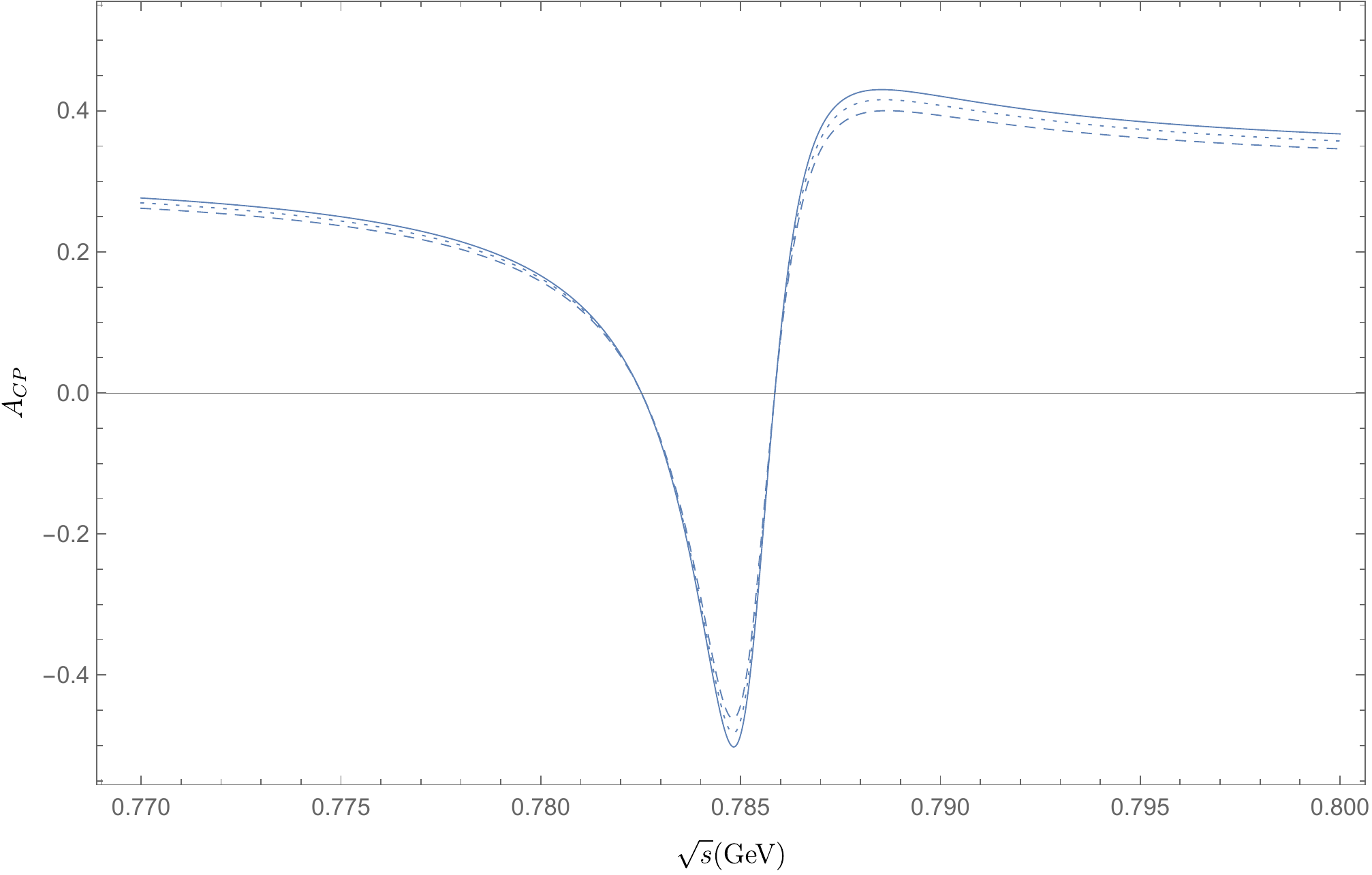}
\caption{\label{Acp plot} The $CP$ violation, $A_{cp}$, as a function of $\sqrt{s}$ for
different CKM matrix elements. The solid line, dot line and dash line corresponds to the maximum, middle, and minimum CKM matrix element for the decay channel of $\bar{B}^0_{s}\rightarrow \rho^0(\omega)K^{*0}\rightarrow\pi^+\pi^-K^{*0}$, respectively.}
\end{figure}

We have investigated the $CP$ violating asymmetry, $A_{CP}$, for the $\bar{B}^0_{s}\rightarrow \rho^0(\omega)K^{*0}\rightarrow\pi^+\pi^-K^{*0}$ of the three-body decay process in the perturbative QCD.  The numerical results of the $CP$ violating asymmetry are shown for the  $\bar{B}^0_{s}\rightarrow \rho^0(\omega)K^{*0}\rightarrow\pi^+\pi^-K^{*0}$ decay process in Fig.~\ref{Acp plot}. It is found that the $CP$ violation can be enhanced via $\rho-\omega$ mixing for the decay channel $\bar{B}^0_{s}\rightarrow \rho^0(\omega)K^{*0}\rightarrow\pi^+\pi^-K^{*0}$ when the invariant mass of $\pi^{+}\pi^{-}$ pair is in the vicinity of the $m_\omega$ resonance within perturbative QCD scheme.

The $CP$ violating asymmetry depends on the weak phase difference $\phi$ from CKM matrix elements and the strong phase difference $\delta$ in the Eq. (\ref{eq:CP-tuidao}). The CKM matrix elements, which relate to $\bar{\rho}$, $A$, $\bar{\eta}$ and $\lambda$, are given in Eq. (\ref{eq: rhobarvalue}). The uncertainties due to the CKM matrix elements are mostly from $\rho$ and $\eta$ since $\lambda$ is well determined. Hence we take the central value of $\lambda=0.226$ in Eq. (\ref{eq: rho value}). In the numerical calculations for the $\bar{B}^0_{s}\rightarrow \rho^0(\omega)K^{*0}\rightarrow\pi^+\pi^-K^{*0}$ decay process, we use $\rho$, $\eta$ and $\lambda=0.226$ vary among the limiting values. The numerical results are shown from Fig.~\ref{Acp plot} with the different parameter values of CKM matrix elements. The solid line, dot line and dash line corresponds to the maximum, middle, and minimum CKM matrix element for the decay channel of $\bar{B}^0_{s}\rightarrow \rho^0(\omega)K^{*0}\rightarrow\pi^+\pi^-K^{*0}$, respectively. We find the numberical results of the $CP$ violation is not sensitive to the CKM matrix elements for the different values of $\rho$ and $\eta$. In Fig.~\ref{Acp plot}, we show the plot of $CP$ violation as a function of $\sqrt{s}$ in the perturbative QCD. From the figure, one can see the $CP$ violation parameter is dependent on $\sqrt{s}$ and changes rapidly by the $\rho-\omega$ mixing  mechanism when the invariant mass of $\pi^{+}\pi^{-}$ pair is in the vicinity of the $m_\omega$ resonance. From the numerical results, it is found that the $CP$ violating asymmetry is large and ranges from -$50.19\%$ to $43.02\%$ via the $\rho-\omega$ mixing mechanism for the process. The maximum $CP$ violating parameter can reach -$48.22^{+1.97}_{-2.04}\%$ for the decay channel of $\bar{B}^0_{s}\rightarrow\pi^+\pi^-K^{*0}$ in the case of ($\rho$, $\eta$). This error corresponds to the CKM parameters.

\begin{figure}
	\centering
\includegraphics[width=0.5\textwidth]{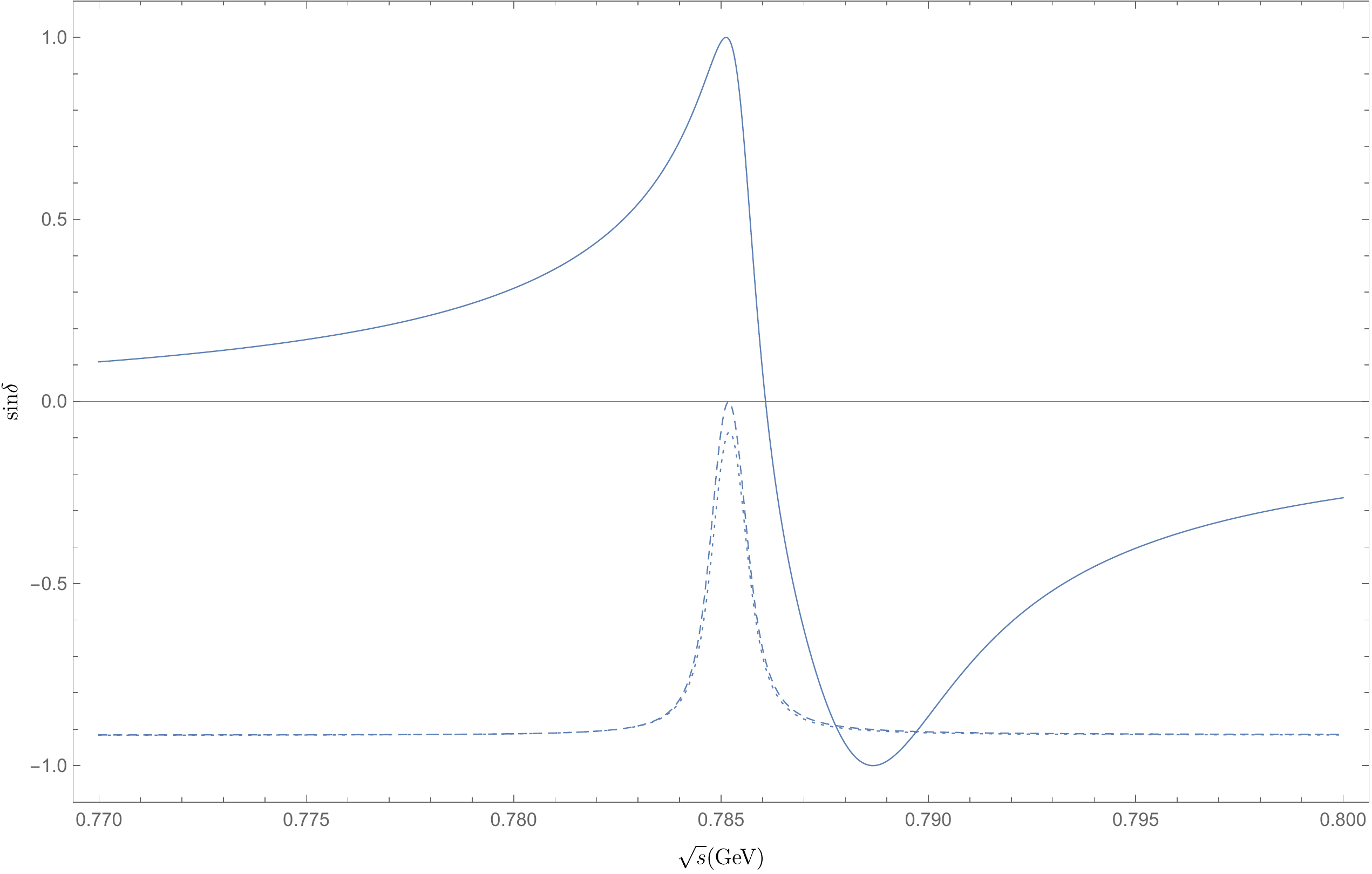}
\caption{\label{sin1 plot} Plot of $\sin\delta$ as a function of $\sqrt{s}$
corresponding to central parameter values of CKM matrix elements
for $\bar{B}^0_{s}\rightarrow \rho^0(\omega)K^{*0}\rightarrow\pi^+\pi^-K^{*0}$.
The solid line, dot line and dash line corresponds to $\sin{\delta_{0}}$, $\sin{\delta_{+}}$ and $\sin{\delta_{-}}$, respectively.}
\end{figure}

\begin{figure}
	\centering
\includegraphics[width=0.5\textwidth]{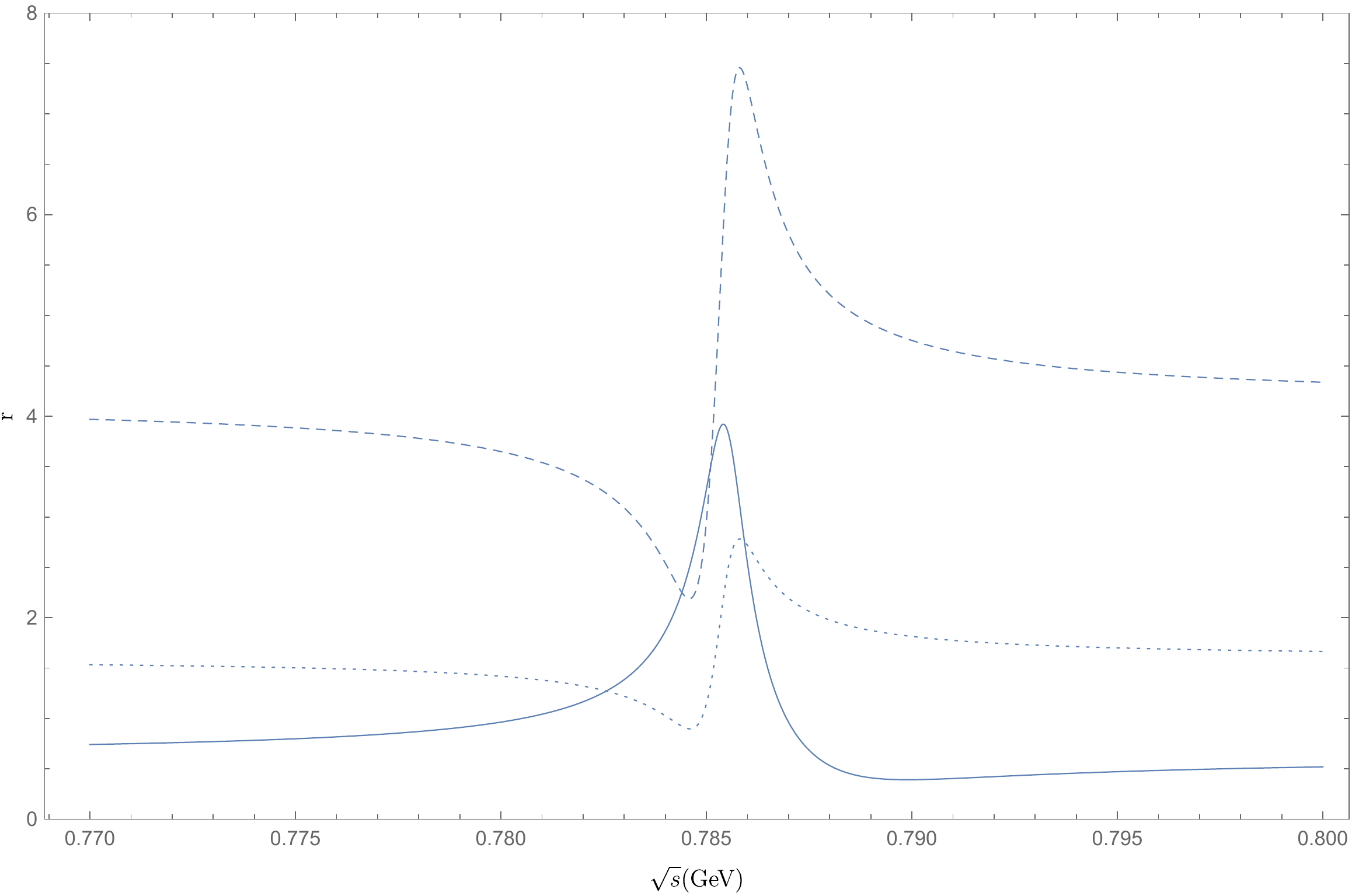}
\caption{\label{r plot} Plot of $r$ as a function of $\sqrt{s}$
corresponding to central parameter values of CKM matrix elements
for $\bar{B}^0_{s}\rightarrow \rho^0(\omega)K^{*0}\rightarrow\pi^+\pi^-K^{*0}$.
The solid line, dot line and dash line corresponds to $r_{0}$, $r_{+}$ and $r_{-}$, respectively.}
\end{figure}

From Eq. (\ref{eq:CP-tuidao}), one can find that the $CP$ violating parameter is related to $r$ and sin$\delta$. 
In Fig.~\ref{sin1 plot} and Fig. \ref{r plot}, we show the plots of
$\sin\delta_0$ ($\sin\delta_+$ and $\sin\delta_-$) and $r_0$ ($r_+$ and $r_-$) as a function of $\sqrt{s}$, respectively.
 We can see that the $\rho-\omega$ mixing mechanism produces a large $\sin\delta_0$ ($\sin\delta_+$ and $\sin\delta_-$) in the vicinity of the $\omega$ resonance. As can be seen from Fig.~\ref{sin1 plot}, the plots vary sharply in the cases of $\sin\delta_0$, $\sin\delta_+$ and $\sin\delta_-$ in the range of the resonance. Meanwhile, $\sin\delta_+$ and $\sin\delta_-$ change weakly compared with the $\sin\delta_0$. It can be seen from Fig. \ref{r plot} that $r_0$, $r_{+}$ and $r_-$ change more rapidly when the $\pi^+ \pi^-$ pairs in the vicinity of the $\omega$ resonance.

We have shown that the $\rho-\omega$ mixing does enhance the direct $CP$ violating asymmetry and provide a mechanism for large $CP$ violation in the perturbative QCD factorization scheme. In other words, it is important to see whether it is possible to observe this large $CP$ violating asymmetry in experiments. This depends on the branching ratio for the decay channel of $\bar{B}^0_{s}\rightarrow \rho^0 K^{*0}$. We will study this problem in the next section.

\subsection{\label{num:BR} Branching ratio via $\rho-\omega$ mixing in $\bar{B}^0_{s}\rightarrow \rho^0(\omega) K^{*0}$}

\begin{figure}
	\centering
	\includegraphics[width=0.5\textwidth]{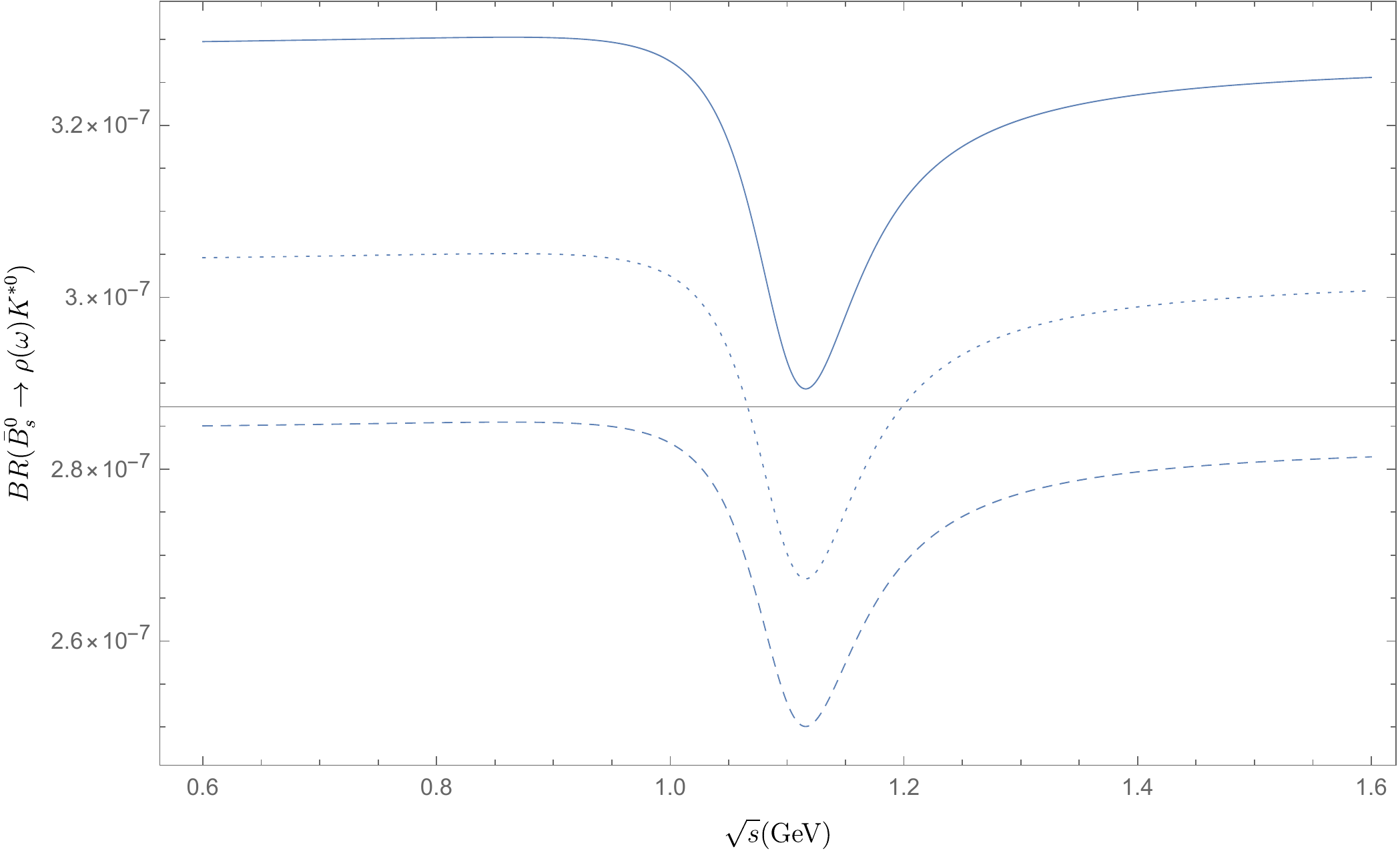}
	\caption{\label{BR plot} The branching ratio, $BR(\bar{B}_s^0\to\rho(\omega) K^{*0})$, as a function of $\sqrt{s}$ for
		different CKM matrix elements. The solid line, dot line and dash line corresponds to the maximum, middle, and minimum CKM matrix element for the decay channel of  $\bar{B}^0_{s}\rightarrow \rho^0(\omega) K^{*0}$, respectively.}
\end{figure}

In the pQCD, we calculate the value of the branching ratio via $\rho-\omega$ mixing mechanism for the decay channel $\bar{B}^0_{s}\rightarrow \rho^0(\omega) K^{*0}$. The numerical result is shown for the decay process in Fig.~\ref{BR plot}. Based on a reasonable parameter range, we obtain the maximum branching ratio of $\bar{B}^0_{s}\rightarrow \rho^0(\omega) K^{*0}$ as ($3.05^{+0.25}_{-0.20}$)$\times 10^{-7}$, which is consistent with the result in \cite{Zou:2015iwa,Yan:2018fif}. The error comes from CKM parameters. On the other hand, although we calculate the branching ratio due to $\rho-\omega$ mixing in the pQCD factorization scheme, we find that the contribution of $\rho-\omega$ mixing to the branching ratio of $\bar{B}^0_{s}\rightarrow \rho^0(\omega) K^{*0}$ is small and can be neglected. However, the $\rho-\omega$ mixing mechanism produces new strong phase differences. This is why the Fig.~\ref{BR plot} presents a tiny effect for the branching ratio of $\bar{B}_s^0 \to \rho^0(\omega) K^{*0}$ when the invariant mass of $\pi^{+}\pi^{-}$ pair is around 1.1 GeV.

The Large Hadron Collider (LHC) is a proton-proton collider that has started at the European Organization for Nuclear Research (CERN). With the designed center-of-mass energy 14 TeV and luminosity $L=10^{34} \text{cm}^{-2} \text{s}^{-1}$, the LHC provides a high energy frontier at TeV-level scale and an opportunity to further improve the consistency test for the CKM matrix. LHCb is a dedicated heavy flavor physics experiments and one of the main projects of LHC. Its main goal is to search for indirect evidence of new physics in $CP$ violation and rare decays in the interactions of beauty and charm hadrons systems, by looking for the effects of new particles in decay processes that are precisely predicted in the SM. Such studies can help us to comprehend the matter-antimatter asymmetry of the universe. Recently, the LHCb collaboration found clear evidence for direct $CP$ violation in some three-body decay channels of $B$ meson. Large $CP$ violation is obtained for the decay channels of $B^{\pm}\rightarrow \pi^{\pm}\pi^{+}\pi^{-}$ in the localized phase spaces region $m^{2}_{\pi^{+}\pi^{-}\text{low}}<0.4$ GeV${^2}$ and $m^{2}_{\pi^{+}\pi^{-}\text{high}}>15$ GeV${^2}$ \cite{Aaij:2013bla,dosReis:2016ayt}. A zoom of the $\pi^+ \pi^-$ invariant mass from the $B^{+}\rightarrow \pi^{+}\pi^{+}\pi^{-}$ decay process is shown  the region  $m^{2}_{\pi^{+}\pi^{-}low}<1$ GeV${^2}$ zone in 
the Ref. \cite{dosReis:2016ayt}. In addition, the branching ratio of $\bar{B}^0_{s}\rightarrow \pi^+\pi^-\phi$ is probed in the $\pi^{+}\pi^{-}$ invariant mass range $400 < m({\pi ^ + }{\pi ^ - }) < 1600$ MeV/$\rm {c^2}$  \cite{Aaij:2016qnm}. In the next years, we expect the LHCb Collaboration to collect date for detecting our prediction of $CP$ violation from the $\bar{B}^0_{s}\rightarrow \rho^0(\omega)K^{*0}\rightarrow\pi^+\pi^-K^{*0}$ decay process when the invariant mass of $\pi^+\pi^-$ is in the vicinity of the $\omega$ resonance.

At the LHC, the $b$-hadrons come from $pp$ collisions. The possible asymmetry between the numbers of the $b$-hadrons $H_b$ and those of their anti-particles $\bar{H}_b$ has been studied by using the intrinsic heavy quark model and the Lund string fragmentation model \cite{Norrbin:1999by,Altarelli:2000ye}. It has been shown that this asymmetry can only reach values of a few percents. In the following discussion, we will ignore this small asymmetry and give the numbers of $H_b \bar{H}_b$ pairs needed for observing our prediction of the $CP$ violating asymmetries. These numbers depend on both the magnitudes of the $CP$ violating asymmetries and the branching ratios of heavy hadron decays which are model dependent. For one-standard-deviation (1$\sigma$) signature and three-standard deviation (3$\sigma$) signature, the numbers of $H_b \bar{H}_b$ pairs we need \cite{Du:1986ai,Eadie:1971qcl,Lyons:1986em}
\begin{eqnarray}
N_{H_b \bar{H}_b}(1\sigma)\sim \frac{1}{BR(\bar{B}_s^0\rightarrow\rho^{0}K^{*0}) A_{CP}^2}(1-A_{CP}^2)  
\end{eqnarray}
and
\begin{eqnarray}
N_{H_b \bar{H}_b}(3\sigma)\sim \frac{9}{BR(\bar{B}_s^0\rightarrow\rho^{0}K^{*0}) A_{CP}^2}(1-A_{CP}^2),  
\end{eqnarray}
where $A_{CP}$ is the $CP$ violation in the process  of $\bar{B}^0_{s}\rightarrow \rho^0(\omega)K^{*0}\rightarrow\pi^+\pi^-K^{*0}$. Now, we can estimate the possibility to observe $CP$ violation. The branching ratio for $\bar{B}_s^0\rightarrow\rho^{0}(\omega) K^{*0}$ is of order $10^{-7}$, then the number $N_{H_b \bar{H}_b}(1\sigma)\sim 10^6$ for $1\sigma$ signature and $10^7$ for $3\sigma$ signature. Theoretically, in order to achieve the current experiments on $b$-hadrons, which can only provide about $10^7$ $B_s \bar{B}_s$ pairs. Therefore, it is very possible to observe the large $CP$ violation for $\bar{B}^0_{s}\rightarrow \rho^0(\omega)K^{*0}\rightarrow\pi^+\pi^-K^{*0}$ when the invariant masses of $\pi^{+}\pi^{-}$ pairs are in the vicinity of the $\omega$ resonance in experiments at the LHC.

\section{\label{sec:conclusion}Summary and conclusion}

In this paper, we have studied the direct $CP$ violation for the decay process of $\bar{B}^0_{s}\rightarrow \rho^0(\omega)K^{*0}\rightarrow\pi^+\pi^-K^{*0}$ in perturbative QCD. It has been found that, by using $\rho-\omega$ mixing, the $CP$ violation can be enhanced at the area of $\omega$ resonance. There is the resonance effect via $\rho-\omega$ mixing which can produce large strong phase in this decay process. As a result, one can find that the maximum $CP$ violation can reach -$50.19\%$ when the invariant mass of the $\pi^+\pi^-$ pair is in the vicinity of the $\omega$ resonance. Furthermore, taking $\rho-\omega$ mixing into account, we have calculated the branching ratio of the decays of $\bar{B}^0_s \rightarrow \rho^0 (\omega) K^{*0}$. We have also given the numbers of  $B_s \bar{B}_s$ pairs required for observing our prediction of the $CP$ violating asymmetries at the LHC experiments. 

In our calculation there are some uncertainties. The major uncertainties come from the input parameters. In particular, these include the CKM matrix element, the particle mass, the perturbative QCD approach and the hadronic parameters (decay constants, the wave functions, the shape parameters and etc). We expect that our predictions will provide useful guidance for future experiments.

\section*{Acknowledgements}
This work (https://arxiv.org/abs/2102.07984) was supported by National Natural Science Foundation of China (Project Numbers 11605041), and the Research Foundation of the young core teacher from Henan province.

\appendix

\section*{\label{Appendix}  Appendix: Related functions defined in the text}

In this appendix we present explicit expressions of the factorizable and non-factorizable amplitudes in Perturbative QCD \cite{Ali:2007ff,Keum:2000ph,Keum:2000wi,Lu:2000em}. The factorizable amplitudes $F_{B_s\to K^*}^{LL,i}(a_i)$, $F_{ann}^{LL,i}(a_i)$ and $F_{ann}^{SP,i}(a_i)$ (i=L,N,T) are written as
\begin{eqnarray}
  f_{M_2} F^{LL,L}_{B_s\to K^*} (a_i)
  &=&8\pi C_FM_{B_s}^4f_{M_2}\int^1_0dx_1dx_3\int^\infty_0b_1db_1b_3db_3 \phi_{B_s}(x_1,b_1) \Big\{a_i(t_a) E_e(t_a) \nonumber\\
  &&\times \left[r_3(1-2x_3)(\phi_{3}^s(x_3)+\phi_3^t(x_3))+(1+x_3)\phi_3(x_3)\right]h_e(x_1,x_3,b_1,b_3)\nonumber\\ 
  &&+ 2r_3\phi_3^s(x_3)a_i(t_a^\prime) E_e(t_a^\prime)h_e(x_3,x_1,b_3,b_1) \Big\}, 
\end{eqnarray}
\begin{eqnarray}
f_{M_2}F^{LL,N}_{B_s\to K^*}(a_i)
&=&8\pi C_FM_{B_s}^4f_{M_2}r_2\int^1_0dx_1dx_3\int^\infty_0b_1db_1b_3db_3\phi_{B_s}(x_1,b_1)\Big\{h_e(x_1,x_3,b_1,b_3) \nonumber\\
&&\times E_e(t_a)a_i(t_a)\left[ 2r_3 \phi_{3}^v(x_3)+r_3 x_3 (\phi_{3}^v(x_3)-\phi_{3}^a(x_3))+ \phi_{3}^T(x_3)\right]  \nonumber\\ 
&&+r_3 \left[ \phi_{3}^v(x_3)+\phi_{3}^a(x_3)\right] E_e(t_a')a_i(t_a')h_e(x_3,x_1,b_3,b_1)\Big\},
\end{eqnarray}
\begin{eqnarray}
f_{M_2}F^{LL,T}_{B_s \to K^*}( a_i)
&=&16\pi C_FM_{B_s}^4f_{M_2}r_2\int^1_0dx_1dx_3\int^\infty_0b_1db_1b_3db_3\phi_{B_s}(x_1,b_1)
  \Big\{h_e(x_1,x_3,b_1,b_3)\nonumber\\
  &&\times \left[ 2r_3 \phi_{3}^v(x_3)-r_3 x_3
  (\phi_{3}^v(x_3)-\phi_{3}^a(x_3))+ \phi_{3}^T(x_3)\right] E_e(t_a)a_i(t_a)
 \nonumber\\
 &&+r_3\left[ \phi_{3}^v(x_3)+\phi_{3}^a(x_3)\right] E_e(t_a')a_i(t_a')h_e(x_3,x_1,b_3,b_1)
 \Big\},
\end{eqnarray}
\begin{eqnarray}
f_{B_s} F_{ann}^{LL,L}( a_i)&=&8\pi
C_FM_{B_s}^4f_{B_s}\int^1_0dx_2dx_3\int^\infty_0b_2db_2b_3db_3\Big\{a_i(t_c)
E_a(t_c) [(x_3-1)\phi_2(x_2)\phi_3(x_3) \nonumber\\
&&-4r_2r_3\phi_2^s(x_2)\phi_3^s(x_3) +2r_2r_3x_3\phi_2^s(x_2)(\phi_3^s(x_3)-\phi_3^t(x_3))] h_a(x_2,1-x_3,b_2,b_3)\nonumber\\
&&+[x_2\phi_2(x_2)\phi_3(x_3)+2r_2r_3(\phi_2^s(x_2)-\phi_2^t(x_2))\phi_3^s(x_3)+2r_2r_3x_2(\phi_2^s(x_2) \nonumber\\
&&+\phi_2^t(x_2))\phi_3^s(x_3)] a_i(t_c^\prime)E_a(t_c^\prime)h_a(1-x_3,x_2,b_3,b_2)\Big\},
\end{eqnarray}
\begin{eqnarray}
f_{B_s}F_{ann}^{LL,N}(a_i)
&=& -8\pi C_FM_{B_s}^4f_{B_s}r_2r_3\int^1_0dx_2dx_3\int^\infty_0b_2db_2b_3db_3\Big\{E_a(t_c)a_i(t_c)h_a(x_2,1-x_3,b_2,b_3))\nonumber\\
&&\times\left[x_3(\phi_2^v(x_2)\phi_3^a(x_3)+\phi_2^a(x_2)\phi_3^v(x_3))+(2-x_3)\left(\phi_2^v(x_2)\phi_3^v(x_3)+\phi_2^a(x_2)\phi_3^a(x_3)\right)\right]\nonumber\\
&&-h_a(1-x_3,x_2,b_3,b_2)[(1+x_2)(\phi_2^a(x_2)\phi_3^a(x_3)+\phi_2^v(x_2)\phi_3^v(x_3))\nonumber\\
&&-(1-x_2)(\phi_2^a(x_2)\phi_3^v(x_3))+\phi_2^v(x_2)\phi_3^a(x_3)]E_a(t_c')a_i(t_c')\Big\},
\end{eqnarray}
\begin{eqnarray}
f_{B_s}F_{ann}^{LL,T}(a_i)
&=& -16\pi C_FM_{B_s}^4f_{B_s}r_2
r_3\int^1_0dx_2dx_3\int^\infty_0b_2db_2b_3db_3 \Big\{\Big[x_3(\phi_2^a(x_2)\phi_3^a(x_3)+\phi_2^v(x_2)\phi_3^v(x_3))\nonumber\\
&&+(2-x_3)(\phi_2^a(x_2)\phi_3^v(x_3)+\phi_2^v(x_2)\phi_3^a(x_3))\Big]
E_a(t_c)a_i(t_c)h_a(x_2,1-x_3,b_2,b_3)\nonumber\\
&&+h_a(1-x_3,x_2,b_3,b_2)[(1-x_2)
(\phi_2^a(x_2)\phi_3^a(x_3)+\phi_2^v(x_2)\phi_3^v(x_3))\nonumber\\
&&
-(1+x_2)(\phi_2^a(x_2)\phi_3^v(x_3)+\phi_2^v(x_2)\phi_3^a(x_3))]
E_a(t_c')a_i(t_c')\Big\},
\end{eqnarray}
\begin{eqnarray}
f_{B_s} F_{ann}^{SP,L}(a_i)&=&16\pi C_FM_{B_s}^4f_{B_s}\int^1_0dx_2dx_3\int^\infty_0b_2db_2b_3db_3\Big\{[(x_3-1)r_3\phi_2(x_2)(\phi_3^s(x_3)+\phi_3^t(x_3))
\nonumber\\
&&-2r_2\phi_2^s(x_2)\phi_3(x_3)] a_i(t_c) E_a(t_c)h_a(x_2,1-x_3,b_2,b_3)-[2r_3\phi_2(x_2)\phi_3^s(x_3) \nonumber\\
&&+r_2x_2(\phi_2^s(x_2)-\phi_2^t(x_2))\phi_3(x_3)] a_i(t_c^\prime)E_a(t_c^\prime)h_a(1-x_3,x_2,b_3,b_2)\Big\}, 
\end{eqnarray}
\begin{eqnarray}
f_{B_s}F_{ann}^{SP,T}(a_i)&=&2
f_{B_s}F_{ann}^{SP,N}(a_i)\nonumber\\
&=&-32\pi
C_FM_{B_s}^4f_{B_s}\int^1_0dx_2dx_3\int^\infty_0b_2db_2b_3db_3\Big\{
r_2(\phi_2^a(x_2)+\phi_2^v(x_2))\phi_3^T(x_3)\nonumber\\
&&\times E_a(t_c)a_i(t_c)h_a(x_2,1-x_3,b_2,b_3)+r_3 \phi_2^T(x_2) \nonumber\\
&&\times (\phi_3^v(x_3)-\phi_3^a(x_3))E_a(t_c')a_i(t_c')h_a(1-x_3,x_2,b_3,b_2)
\Big\},
\end{eqnarray}
with the color factor ${C_F} = 3/4$, $a_i$ represents the corresponding Wilson coefficients for specific decay
channels and  $f_{M_2}$, $f_{B_s}$ refer to the decay constants of $M_2$ ($\rho$ or $\omega)$ and $\bar{B}_{s}^0$ mesons. In the above functions, $r_{2} (r_{3}) = M_{V_2}(M_{V_3}) /M_{B_s}$ and $\phi_{2} (\phi_{3}) = \phi_{\rho/\omega}(\phi_{K^{*}})$,
with $M_{B_s}$ and $M_{V_2}(m_{V_3})$ being the masses of the
initial and final states.

The non-factorizable amplitudes $M_{B_s \to K^*}^{LL,i}(a_i)$, $M_{B_s \to K^*}^{LR,i}(a_i)$, $M_{B_s \to K^*}^{SP,i}(a_i)$, $M_{ann}^{LL,i}(a_i)$ and $M_{ann}^{LR,i}(a_i)$(i=L,N,T) are written as
\begin{eqnarray}
 M_{B_s\to K^*}^{LL,L}(a_i)&=&32\pi
C_FM_{B_s}^4/\sqrt{6}\int^1_0dx_1dx_2dx_3\int^\infty_0b_1db_1b_2db_2
\phi_{B_s}(x_1,b_1)\phi_2(x_2)
\nonumber\\
&&\times
\Big\{\Big[(1-x_2)\phi_3(x_3)-r_3x_3(\phi_3^s(x_3)-\phi_3^t(x_3))\Big]
a_i(t_b)E_e^\prime(t_b)\nonumber\\
&&\times h_n(x_1,1-x_2,x_3,b_1,b_2)+h_n(x_1,x_2,x_3,b_1,b_2)\nonumber\\
&&\times\Big[r_3x_3(\phi_3^s(x_3)+\phi_3^t(x_3))-(x_2+x_3)\phi_3(x_3)\Big] a_i(t_b^\prime) E_e^\prime(t_b^\prime)\Big\}, 
 \end{eqnarray}
\begin{eqnarray}
M_{B_s \to K^{*}}^{LL,N}(a_i)&=&32\pi
C_FM_{B_s}^4r_2/ \sqrt{6}\int^1_0dx_1dx_2dx_3\int^\infty_0b_1db_1b_2db_2\phi_{B_s}(x_1,b_1) \nonumber\\
&& \times \Big\{\left[x_2(\phi_2^v(x_2)+\phi_2^a(x_2))\phi_3^T(x_3) -2r_3(x_2+x_3)(\phi_2^a(x_2)\phi_3^a(x_3)+\phi_2^v(x_2)\phi_3^v(x_3))\right] \nonumber\\
&&\times h_n(x_1,x_2,x_3,b_1,b_2)E_e'(t_b')a_i(t_b')+(1-x_2)(\phi_2^v(x_2)+\phi_2^a(x_2))\phi_3^T(x_3)\nonumber\\
&&\times E_e'(t_b)a_i(t_b) h_n(x_1,1-x_2,x_3,b_1,b_2)\Big\},
\end{eqnarray}
\begin{eqnarray}
M_{B_s\to K^{*}}^{LL,T}(a_i)&=&64\pi C_FM_{B_s}^4r_2/
\sqrt{6}\int^1_0dx_1dx_2dx_3\int^\infty_0b_1db_1b_2db_2\phi_{B_s}(x_1,b_1)\Big\{
E_e'(t_b')a_i(t_b')\nonumber \\
&&\times \big[x_2(\phi_2^v(x_2)+\phi_2^a(x_2))\phi_3^T(x_3)
-2r_3(x_2+x_3)(\phi_2^v(x_2)\phi_3^a(x_3)\nonumber \\
&&+\phi_2^a(x_2)\phi_3^v(x_3))\big]h_n(x_1,x_2,x_3,b_1,b_2)+(1-x_2)[\phi_2^v(x_2)+\phi_2^a(x_2)]\phi_3^T(x_3)\nonumber \\
&&\times  E_e'(t_b)a_i(t_b) h_n(x_1,1-x_2,x_3,b_1,b_2)\Big\},
\end{eqnarray}
\begin{eqnarray} M_{B_s\to K^{*}}^{LR,L}( a_i)&=&32\pi C_FM_{B_s}^4r_2/\sqrt{6}
\int^1_0dx_1dx_2dx_3\int^\infty_0b_1db_1b_2db_2\phi_{B_s}(x_1,b_1)\nonumber\\
&&\times \Big\{h_n(x_1,1-x_2,x_3,b_1,b_2)\Big[(1-x_2)\phi_3(x_3)
\left(\phi_2^s(x_2)+\phi_2^t(x_2)\right)\nonumber\\
&&+r_3x_3\left(\phi_2^s(x_2)-\phi_2^t(x_2)\right)
\left(\phi_3^s(x_3)+\phi_3^t(x_3)\right)\nonumber\\
&&+(1-x_2)r_3\left(\phi_2^s(x_2)+\phi_2^t(x_2)\right)\left(\phi_3^s(x_3)
-\phi_3^t(x_3)\right)\Big]a_i(t_b)
E_e^\prime(t_b) \nonumber\\
&&-h_n(x_1,x_2,x_3,b_1,b_2)\Big[x_2\phi_3(x_3)(\phi_2^s(x_2)-\phi_2^t(x_2))\nonumber\\
&&+r_3x_2(\phi_2^s(x_2)-\phi_2^t(x_2))(\phi_3^s(x_3)-\phi_3^t(x_3))\nonumber\\
&&+r_3x_3(\phi_2^s(x_2)+\phi_2^t(x_2))(\phi_3^s(x_3)+\phi_3^t(x_3))\Big]a_i(t^\prime_b)
E_e^\prime(t_b^\prime)\Big\}, 
\end{eqnarray}
\begin{eqnarray}
M_{B_s\to K^{*}}^{LR,T}(a_i)&=&
2M_{B_s\to K^{*}}^{LR,N}(a_i)\nonumber\\
&=&64\pi C_FM_{B_s}^4/\sqrt{6}\int^1_0dx_1dx_2dx_3\int^\infty_0b_1db_1b_2db_2\phi_{B_s}(x_1,b_1)\nonumber
\\
&&\times r_3 x_3 \phi_2^T(x_2)(\phi_3^v(x_3)-\phi_3^a(x_3))\nonumber\\
&& \times \Big\{E_e'(t_b)a_i(t_b)
h_n(x_1,1-x_2,x_3,b_1,b_2)+E_e'(t_b')a_i(t_b')
h_n(x_1,x_2,x_3,b_1,b_2)\Big\},
\end{eqnarray}
\begin{eqnarray} M^{SP,L}_{B_s\to K^{*}}( a_i) &=&32\pi C_F
M_{B_s}^4/\sqrt{6}\int^1_0dx_1dx_2dx_3\int^\infty_0b_1db_1b_2db_2
\phi_{B_s}(x_1,b_1)\phi_2(x_2)
\nonumber\\
&&\times\Big\{
\Big[(x_2-x_3-1)\phi_3(x_3)+r_3x_3(\phi_3^s(x_3)+\phi_3^t(x_3))\Big]\nonumber\\
&&\times
a_i(t_b)E_e^\prime(t_b)h_n(x_1,1-x_2,x_3,b_1,b_2)+a_i(t_b^\prime)
E^\prime_e(t_b^\prime)\nonumber\\
&&\times
 \Big[x_2\phi_3(x_3)+r_3x_3(\phi_3^t(x_3)-\phi_3^s(x_3))\Big]h_n(x_1,x_2,x_3,b_1,b_2)\Big\},
\end{eqnarray}
\begin{eqnarray} M^{SP,N}_{B_s\to K^{*}}(a_i) &=&32 \pi C_F
M_{B_s}^4/\sqrt{6}\int^1_0dx_1dx_2dx_3\int^\infty_0b_1db_1b_2db_2
\phi_{B_s}(x_1,b_1)r_2\nonumber \\
&&\times\Big\{ x_2
(\phi_2^v(x_2)-\phi_2^a(x_2))\phi_3^T(x_3)E_e'(t_b')a_i(t_b')
h_n(x_1,x_2,x_3,b_1,b_2)\nonumber\\
&&+h_n(x_1,1-x_2,x_3,b_1,b_2)[(1-x_2)(\phi_2^v(x_2)-\phi_2^a(x_2))\phi_3^T(x_3)\nonumber \\
&&-2r_3(1-x_2+x_3)
(\phi_2^v(x_2)\phi_3^v(x_3)-\phi_2^a(x_2)\phi_3^a(x_3))]E_e'(t_b)a_i(t_b)
\Big\} ,
 \end{eqnarray}
\begin{eqnarray}
 M^{SP,T}_{B_s\to K^{*}}(a_i) &=&64\pi C_F
M_{B_s}^4/\sqrt{6}\int^1_0dx_1dx_2dx_3\int^\infty_0b_1db_1b_2db_2\phi_{B_s}(x_1,b_1)r_2 \nonumber \\
&&\times\Big\{ x_2
(\phi_2^v(x_2)-\phi_2^a(x_2))\phi_3^T(x_3)E_e'(t_b')a_i(t_b')
h_n(x_1,x_2,x_3,b_1,b_2)\nonumber
\\
&& +h_n(x_1,1-x_2,x_3,b_1,b_2)[(1-x_2)(\phi_2^v(x_2)-\phi_2^a(x_2))\phi_3^T(x_3)\nonumber \\
&& -2r_3(1-x_2+x_3)
(\phi_2^v(x_2)\phi_3^a(x_3)-\phi_2^a(x_2)\phi_3^v(x_3))]E_e'(t_b)a_i(t_b)
\Big\},
\end{eqnarray}
\begin{eqnarray}
M_{ann}^{LL,L}( a_i)&=&32\pi C_FM_{B_s}^4/\sqrt
{6}\int^1_0dx_1dx_2dx_3\int^\infty_0b_1db_2b_2db_2\phi_{B_s}(x_1,b_1)\Big\{h_{na}(x_1,x_2,x_3,b_1,b_2)\nonumber\\
&&\times \Big[r_2r_3x_3(\phi_2^s(x_2)-\phi_2^t(x_2))(\phi_3^s(x_3)+\phi_3^t(x_3))-(x_2\phi_2(x_2)\phi_3(x_3)+4r_2r_3
\phi_2^s(x_2)\phi_3^s(x_3))
\nonumber\\
&& +r_2r_3(1-x_2)(\phi_2^s(x_2)+\phi_2^t(x_2))(\phi_3^s(x_3)-\phi_3^t(x_3)) \Big]a_i(t_d)
E_a^\prime(t_d)+h_{na}^\prime(x_1,x_2,x_3,b_1,b_2)\nonumber\\
&& \times \Big[(1-x_3)\phi_2(x_2)\phi_3(x_3) +(1-x_3)r_2r_3(\phi_2^s(x_2)+\phi_2^t(x_2))(\phi_3^s(x_3)-\phi_3^t(x_3))
\nonumber\\
&& +x_2r_2r_3(\phi_2^s(x_2)-\phi_2^t(x_2))(\phi_3^s(x_3)+\phi_3^t(x_3))\Big]
a_i(t_d^\prime)
E_a^\prime(t_d^\prime)\Big\},
\end{eqnarray}
\begin{eqnarray}
M_{ann}^{LL,N}(a_i)
&=&-64\pi C_FM_{B_s}^4r_2 r_3/\sqrt
{6}\int^1_0dx_1dx_2dx_3\int^\infty_0b_1db_2b_2db_2\phi_{B_s}(x_1,b_1)[\phi_2^v(x_2)\phi_3^v(x_3)\nonumber\\
&& +\phi_2^a(x_2)\phi_3^a(x_3)]
E_a'(t_d)a_i(t_d)h_{na}(x_1,x_2,x_3,b_1,b_2),
\end{eqnarray}
\begin{eqnarray}
M_{ann}^{LL,T}(a_i)
&=&-128 \pi C_FM_{B_s}^4r_2 r_3/\sqrt
{6}\int^1_0dx_1dx_2dx_3\int^\infty_0b_1db_2b_2db_2\phi_{B_s}(x_1,b_1)[\phi_2^v(x_2)\phi_3^a(x_3)\nonumber\\
&&+\phi_2^a(x_2)\phi_3^v(x_3)]
E_a'(t_d)a_i(t_d)h_{na}(x_1,x_2,x_3,b_1,b_2),
\end{eqnarray}
\begin{eqnarray}
M_{ann}^{LR,L}(a_i)&=&32\pi C_FM_{B_s}^4/\sqrt
{6}\int^1_0dx_1dx_2dx_3\int^\infty b_1db_1b_2db_2\phi_{B_s}(x_1,b_1)\Big\{h_{na}(x_1,x_2,x_3,b_1,b_2)\nonumber\\
&&\times \Big[r_2(x_2-2)(\phi_2^s(x_2)+\phi_2^t(x_2))
\phi_3(x_3)+r_3(1+x_3)\phi_2(x_2)(\phi_3^s(x_3)-\phi_3^t(x_3))\Big]a_i(t_d)E_a^\prime(t_d)
\nonumber\\
&&+h_{na}^\prime
(x_1,x_2,x_3,b_1,b_2)\Big[-r_2x_2\left(\phi_2^s(x_2)+\phi_2^t(x_2)\right)\phi_3(x_3)
\nonumber\\
&&+r_3(1-x_3)\phi_2(x_2)
(\phi_3^s(x_3)-\phi_3^t(x_3))\Big]  a_i(t_d^\prime)E_a^\prime(t_d^\prime)
\Big\}, 
\end{eqnarray}
\begin{eqnarray}
M_{ann}^{LR,T}(a_i)&=&2  M_{ann}^{LR,N}(a_i)\nonumber\\
&=&-64\pi C_FM_{B_s}^4 /\sqrt {6}\int^1_0dx_1dx_2dx_3\int^\infty
b_1db_1b_2db_2\phi_{B_s}(x_1,b_1)
\Big\{h_{na}'(x_1,x_2,x_3,b_1,b_2) \nonumber\\
&&\times \left[r_2x_2
(\phi_2^v(x_2)+\phi_2^a(x_2))\phi_3^T(x_3)-r_3(1-x_3)\phi_2^T(x_2)
(\phi_3^v(x_3)-\phi_3^a(x_3))\right]
E_a'(t_d')a_i(t_d')\nonumber\\
&& +\left[r_2(2-x_2)(\phi_2^v(x_2)+\phi_2^a(x_2))\phi_3^T(x_3)
-r_3(1+x_3)\phi_2^T(x_2)(\phi_3^v(x_3)-\phi_3^a(x_3))\right]\nonumber\\
&& \times E_a'(t_d)a_i(t_d)h_{na}(x_1,x_2,x_3,b_1,b_2)\Big\}.
\end{eqnarray}

The hard scale t are chosen as the maximum of the virtuality of the internal momentum transition in the hard amplitudes, including $1/b_i$:
\begin{eqnarray}
t_a&=&\mbox{max}\{{\sqrt
{x_3}M_{B_s},1/b_1,1/b_3}\},\\
t_a^\prime&=&\mbox{max}\{{\sqrt
{x_1}M_{B_s},1/b_1,1/b_3}\},\\
t_b&=&\mbox{max}\{\sqrt
{x_1x_3}M_{B_s},\sqrt{|1-x_1-x_2|x_3}M_{B_s},1/b_1,1/b_2\},\\
t_b^\prime&=&\mbox{max}\{\sqrt{x_1x_3}M_{B_s},\sqrt
{|x_1-x_2|x_3}M_{B_s},1/b_1,1/b_2\},\\
t_c&=&\mbox{max}\{\sqrt{1-x_3}M_{B_s},1/b_2,1/b_3\},\\
t_c^\prime
&=&\mbox{max}\{\sqrt {x_2}M_{B_s},1/b_2,1/b_3\},\\
t_d&=&\mbox{max}\{\sqrt {x_2(1-x_3)}M_{B_s},
\sqrt{1-(1-x_1-x_2)x_3}M_{B_s},1/b_1,1/b_2\},\\
t_d^\prime&=&\mbox{max}\{\sqrt{x_2(1-x_3)}M_{B_s},\sqrt{|x_1-x_2|(1-x_3)}M_{B_s},1/b_1,1/b_2\}.
\end{eqnarray}

The function h, coming from the Fourier transform of hard part H, are written as
\begin{eqnarray}
h_e(x_1,x_3,b_1,b_3)&=&\left[\theta(b_1-b_3)I_0(\sqrt
x_3M_{B_s}b_3)K_0(\sqrt
x_3 M_{B_s}b_1)\right.\nonumber\\
&& \left.+\theta(b_3-b_1)I_0(\sqrt x_3M_{B_s}b_1)K_0(\sqrt
x_3M_{B_s}b_3)\right]K_0(\sqrt {x_1x_3}M_{B_s}b_1)S_t(x_3),
\end{eqnarray}
\begin{eqnarray}
h_n(x_1,x_2,x_3,b_1,b_2)&=&\left[\theta(b_2-b_1)K_0(\sqrt
{x_1x_3}M_{B_s}b_2)I_0(\sqrt
{x_1x_3}M_{B_s}b_1)\right. \nonumber\\
&&\;\;\;\left. +\theta(b_1-b_2)K_0(\sqrt
{x_1x_3}M_{B_s}b_1)I_0(\sqrt{x_1x_3}M_{B_s}b_2)\right]\nonumber\\
&&\times
\left\{\begin{array}{ll}\frac{i\pi}{2}H_0^{(1)}(\sqrt{(x_2-x_1)x_3}
M_{B_s}b_2),& x_1-x_2<0\\
K_0(\sqrt{(x_1-x_2)x_3}M_{B_s}b_2),& x_1-x_2>0
\end{array}
\right. ,
\end{eqnarray}

\begin{eqnarray}
h_a(x_2,x_3,b_2,b_3)&=&(\frac{i\pi}{2})^2
S_t(x_3)\Big[\theta(b_2-b_3)H_0^{(1)}(\sqrt{x_3}M_{B_s}b_2)J_0(\sqrt
{x_3}M_{B_s}b_3)\nonumber\\
&&\;\;+\theta(b_3-b_2)H_0^{(1)}(\sqrt {x_3}M_{B_s}b_3)J_0(\sqrt
{x_3}M_{B_s}b_2)\Big]H_0^{(1)}(\sqrt{x_2x_3}M_{B_s}b_2),
\end{eqnarray}
\begin{eqnarray}
h_{na}(x_1,x_2,x_3,b_1,b_2)&=&\frac{i\pi}{2}\left[\theta(b_1-b_2)H^{(1)}_0(\sqrt
{x_2(1-x_3)}M_{B_s}b_1)J_0(\sqrt {x_2(1-x_3)}M_{B_s}b_2)\right. \nonumber\\
&&\;\;\left.
+\theta(b_2-b_1)H^{(1)}_0(\sqrt{x_2(1-x_3)}M_{B_s}b_2)J_0(\sqrt
{x_2(1-x_3)}M_{B_s}b_1)\right]\nonumber\\
&&\;\;\;\times K_0(\sqrt{1-(1-x_1-x_2)x_3}M_{B_s}b_1),
\end{eqnarray}
\begin{eqnarray}
h_{na}^\prime(x_1,x_2,x_3,b_1,b_2)&=&\frac{i\pi}{2}\left[\theta(b_1-b_2)H^{(1)}_0(\sqrt
{x_2(1-x_3)}M_{B_s}b_1)J_0(\sqrt{x_2(1-x_3)}M_{B_s}b_2)\right. \nonumber\\
&&\;\;\;\left. +\theta(b_2-b_1)H^{(1)}_0(\sqrt
{x_2(1-x_3)}M_{B_s}b_2)J_0(\sqrt{x_2(1-x_3)}M_{B_s}b_1)\right]\nonumber\\
&&\;\;\;\times
\left\{\begin{array}{ll}\frac{i\pi}{2}H^{(1)}_0(\sqrt{(x_2-x_1)(1-x_3)}M_{B_s}b_1),&
x_1-x_2<0\\
K_0(\sqrt {(x_1-x_2)(1-x_3)}M_{B_s}b_1),&
x_1-x_2>0\end{array}\right. ,
\end{eqnarray}
where $J_0$ and ${Y}_0$ are the Bessel function with $H_0^{(1)}(z) = \mathrm{J}_0(z) + i\, \mathrm{Y}_0(z)$.

The threshold re-sums factor $S_t$ follows the parameterized  
\begin{eqnarray}
S_t(x)=\frac{2^{1+2c}\Gamma(3/2+c)}{\sqrt \pi \Gamma(1+c)}[x(1-x)]^c,
\end{eqnarray}
where the parameter $c=0.4$.

The evolution factors $E^{(\prime)}_e$ and $E^{(\prime)}_a$ entering in the expressions for the matrix elements are given by
\begin{eqnarray}
E_e(t)&=&\alpha_s(t) \exp[-S_B(t)-S_{M_3}(t)], \ \ \ \ E'_e(t)=\alpha_s(t) \exp[-S_B(t)-S_{M_2}(t)-S_{M_3}(t)]|_{b_1=b_3},\\
E_a(t)&=&\alpha_s(t) \exp[-S_{M_2}(t)-S_{M_3}(t)],\ \ \ \ E'_a(t)=\alpha_s(t) \exp[-S_B(t)-S_{M_2}(t)-S_{M_3}(t)]|_{b_2=b_3},
\end{eqnarray}
in which the Sudakov exponents are defined as
\begin{eqnarray}
S_B(t)&=&s\left(x_1\frac{M_{B_s}}{\sqrt2},b_1\right)+\frac{5}{3}\int^t_{1/b_1}\frac{d\bar \mu}{\bar\mu}\gamma_q(\alpha_s(\bar \mu)),\\
S_{M_2}(t)&=&s\left(x_2\frac{M_{B_s}}{\sqrt2},b_2\right)+s\left((1-x_2)\frac{M_{B_s}}{\sqrt2},b_2\right)+2\int^t_{1/b_2}\frac{d\bar \mu}{\bar
\mu}\gamma_q(\alpha_s(\bar \mu)), \\
S_{M_3}(t)&=&s\left(x_3\frac{M_{B_s}}{\sqrt2},b_3\right)+s\left((1-x_3)\frac{M_{B_s}}{\sqrt2},b_3\right)+2\int^t_{1/b_3}\frac{d\bar \mu}{\bar
	\mu}\gamma_q(\alpha_s(\bar \mu)),
\end{eqnarray}
where $\gamma_q=-\alpha_s/\pi$ is the anomalous dimension of the quark. The explicit form for the  function $s(Q,b)$ is:
\begin{eqnarray}
s(Q,b)&=&\frac{A^{(1)}}{2\beta_{1}}\hat{q}\ln\left(\frac{\hat{q}}{\hat{b}}\right)-\frac{A^{(1)}}{2\beta_{1}}\left(\hat{q}-\hat{b}\right)+
\frac{A^{(2)}}{4\beta_{1}^{2}}\left(\frac{\hat{q}}{\hat{b}}-1\right)
-\left[\frac{A^{(2)}}{4\beta_{1}^{2}}-\frac{A^{(1)}}{4\beta_{1}}
\ln\left(\frac{e^{2\gamma_E-1}}{2}\right)\right]
\ln\left(\frac{\hat{q}}{\hat{b}}\right)
\nonumber \\
&&+\frac{A^{(1)}\beta_{2}}{4\beta_{1}^{3}}\hat{q}\left[
\frac{\ln(2\hat{q})+1}{\hat{q}}-\frac{\ln(2\hat{b})+1}{\hat{b}}\right]
+\frac{A^{(1)}\beta_{2}}{8\beta_{1}^{3}}\left[
\ln^{2}(2\hat{q})-\ln^{2}(2\hat{b})\right],
\end{eqnarray}
where the variables are defined by
\begin{eqnarray}
\hat q\equiv \mbox{ln}[Q/(\sqrt 2\Lambda)],~~~ \hat b\equiv
\mbox{ln}[1/(b\Lambda)], 
\end{eqnarray}
 and the coefficients
$A^{(i)}$ and $\beta_i$ are 
\begin{eqnarray}
\beta_1=\frac{33-2n_f}{12},~~\beta_2=\frac{153-19n_f}{24},\nonumber\\
A^{(1)}=\frac{4}{3},~~A^{(2)}=\frac{67}{9}
-\frac{\pi^2}{3}-\frac{10}{27}n_f+\frac{8}{3}\beta_1\mbox{ln}(\frac{1}{2}e^{\gamma_E}),
\end{eqnarray}
with $n_f$ is the number of the quark flavors and $\gamma_E$ is the
Euler constant.

\end{spacing}
\end{document}